\documentclass[twocolumn,showpacs,preprintnumbers,amsmath,amssymb]{revtex4}

\ifx\pdfoutput\undefined
\usepackage[dvips]{graphicx}
\usepackage[dvipdfm,
            pdfstartview=FitH,
            CJKbookmarks=true,
            bookmarksnumbered=true,
            bookmarksopen=true,
            colorlinks=true, 
            pdfborder=001,   
            citecolor=blue,%
            linkcolor=blue
            ]{hyperref}
\AtBeginDvi{} 
\else
\usepackage[dvipdf,
            pdfstartview=FitH,
            CJKbookmarks=true,
            bookmarksnumbered=true,
            bookmarksopen=true,
            colorlinks=true,   
            citecolor=blue,%
            linkcolor=blue       　
            ]{hyperref}
\usepackage[pdftex]{graphicx}
\fi

\usepackage{caption2}
\usepackage{graphicx}
\usepackage{dcolumn}
\usepackage{bm}
\usepackage{subfigure}
\usepackage{subfigure}
\usepackage{picinpar}
\usepackage{picins}
\usepackage{epic}
\usepackage{caption2}
\usepackage{amsmath}
\usepackage{amsthm}
\usepackage{latexsym}
\usepackage{amssymb}
\usepackage{epsfig}
\begin{document}

\preprint{preprint}

\title{A Cellular Automata Model with Probability Infection and Spatial Dispersion}
\author{Jin Zhen}%
 \email{jinzhn@263.net}
\author{Liu Quanxing}
 \email{liuqx315@sina.com}
\affiliation{%
Department of mathematics, North University of China,\\
Taiyuan, Shan'xi, 030051, People's Republic of China
}%

\author{Mainul Haque}
\email{mainul.haque@rediffmail.com}
\affiliation{
Department of Mathematics, Krishnath college, \\
Berhampore, Mursidabad, West Bengal, India-742101.
}%

\date{\today}

\begin{abstract}
In this article, we have proposed an epidemic model by using
probability cellular automata theory. The essential mathematical
features are analyzed with the help of stability theory. We have
given an alternative modelling approach for the spatiotemporal
system which is more realistic and satisfactory from the practical
point of view. A discrete and spatiotemporal approach are shown by
using cellular automata theory. It is interesting to note that
both size of the endemic equilibrium and density of the individual
increase with the increasing of the neighborhood size and
infection rate, but the infections decrease with the increasing of
the recovery rate. The stability of the system around the positive
interior equilibrium have been shown by using suitable Lyapunov
function. Finally experimental data simulation for SARS disease in
China and a brief discussion conclude the paper.
\end{abstract}

\maketitle %
\section{Introduction}
The foundation of modern mathematical epidemiology based on the
compartment models were laid in the early 20th century. From the
middle 20th century, mathematical epidemiology has been growing
exponentially. Various mathematical models have been proposed and
analyzed. Although epidemic models are studied too much from that
time but little attention has been paid on the localized processes
of pathogen transmission between susceptible, exposed and infected
individuals. In the epidemic model, generally, it is assumed that
the population is a continuous entity and as a result, it is often
neglected  that the population are composed of single interacting
individuals. In fact, the spread of disease must be governed by
the localized process~\cite{Knu1}. The cellular automata (CA) or
lattice gas cellular automata (LGCA) is a discrete approach to the
time, spatial and state. It shows the features of the epidemic
model clearly with mathematical analysis and numerical simulations
in a very simple way~\cite{Knu14}. In a class of spatial epidemic
models, it is supposed that the lattice of habitable sites, where
no more than one individual can occupy any particular site.
Spatially structured pathogen transmission says  that the
probability of susceptible host acquires some disease depends more
on the local density of the exposed and infected host than that of
the global~\cite{Knu2}. However, in most of these models, local
host density is treated as an equivalent to the global host
density, since every lattice site is occupied with equal
probability. The following spatial heterogeneity is generated by
exogenously, $H$. The varieties of the local host densities are
very important to understand the role of spatial heterogeneity in
the plants and animals communities~\cite{Knu3,Knu4}.

Harcourt~\cite{Knu5} and Kobayashi~\cite{Knu6} have studied an
epidemic model on the Citrus Variegated Chlorosis (CVC) by
considering spatial dispersion and obtained a interesting negative
binomial distribution.  Duryea~\cite{Knu7} has studied an SIS
epidemic model with varying pupolation density and obtained the
condition for which the pathogen extinct.  Similar results have
been shown by Mollision~\cite{Knu2} and Ahmed and
Agiza~\cite{Knu15}. Keeling~\cite{keel} have studied the effects
of local spatial structure on epidemiological invasions in an
epidemic model. Here we have considered an SEIS epidemic model
with infectious force in the exposed and infected period, like as
in~\cite{Knu17,Knu18}. In this SEIS model we have considered the
spatial dispersion and obtained the MF transmission threshold for
epidemic persistence, $\mathcal{R}_c$, by using probability
cellular automata. We have also shown the nature of solution of
the system around the trivial equilibrium (0,0) by using Lyapunov
function. Finally numerical simulations have been performed by the
help of experimental data for the SARS disease in China.

\section{Analysis for the model} In order to get the same neighborhood for
each site of the boundary, we treat the plane as a two-dimensional
torus with $J$ sites, where $J\gg 1$. The constant $J$ is the
carrying capacity of the environment, which is usually determined
by the availability of the resources. In more general cases, the
neighborhood for each site $k$ are shown in \autoref{myfigure1}
and \autoref{myfigure2}. The \autoref{myfigure3} shows the same
fact with a special black point neighborhood for site $k$. Let us
suppose that the individual dispersion in the sites of the
two-dimensional torus for the global state set is dented by $Z(t)$
at time $t$. Then we have
$$Z(t)=\{z_1(t), z_2(t),\cdots,z_{J}(t)\}.$$ In the above global set,
the $z_{k}(t)$ denotes the k-th set of local states at time $t$.
Each site of $J$ is empty or occupied by at most one individual.
So, in the epidemic each site state belongs to the set
$\mathcal{S}$, where $\mathcal{S}=\{\rm
susceptible,exposed,infected,empty\}$. Hence, each $z_{k}(t)$
belongs to a finite set of elementary states. The recoveree from
pathogen infection does not confer immunity, so the recovered
individuals will again be treated as susceptible individual.

\par
 Let us assume that there exist $N$ individual habitable
sites ( or cells) at time $t$ and is denoted by $N(t)$. Each of
the $N$ site contains a single host, where no more than one
individual can occupy any particular site. Then there are $J-N(t)$
number of  empty sites at time $t$, where the empty site
represents no individual host is there. We denote $H(t)$ as the
global density for the individuals at time $t$. Since $J$ is the
maximum number of individual for the system. So we
have\begin{equation}\label{EQ0}
 H(t)=\frac{N(t)}{J},\hspace{1cm}(H(t)\leq 1).
\end{equation}

 Note that in the Eq.~(\ref{EQ0}), $J>N(t)$, therefore we
get $H(t)\leq1$. In this model, we assume that the natural birth
rate equal to it's natural death rate and $d$ is the mortality
rate due to the infection.
\par
\begin{minipage}[b]{0.45\textwidth}
\centering
\begin{picture}(110,120)
\put(-20,5){\grid(105,105)(15,15)} \put(30,54){$k$}
\put(18,57){\circle*{5}} \put(50,57){\circle*{5}}
\put(33,43){\circle*{5}} \put(33,72){\circle*{5}}
\end{picture}
\centering \caption{Von Neumann Neighborhood.}\label{myfigure1}
\end{minipage}
\begin{minipage}[b]{0.45\textwidth}
\centering
\begin{picture}(200,120)
\put(50,5){\grid(105,105)(15,15)} \put(102,54){$k$}
\put(88,72){\circle*{5}} \put(88,57){\circle*{5}}
\put(88,43){\circle*{5}} \put(120,72){\circle*{5}}
\put(120,43){\circle*{5}} \put(120,57){\circle*{5}}
\put(103,43){\circle*{5}} \put(103,72){\circle*{5}}
\end{picture}
\centering \caption{Moore Neighborhood.}\label{myfigure2}
\end{minipage}

\begin{minipage}[b]{0.45\textwidth}
\centering
\begin{picture}(110,75)
\put(-10,5){\grid(105,45)(15,15)} \put(40,24){$k$}
\put(30,27){\circle*{5}} \put(60,27){\circle*{5}}
\end{picture}
\centering \caption{ $r=1/2$.}\label{myfigure3}
\end{minipage}
\setcaptionwidth{5cm}
\begin{minipage}[b]{0.45\textwidth}
\centering
\begin{picture}(100,75)
 \put(0,20){\framebox(20,15){\textmd{$S$}}}
 \put(50,20){\framebox(20,15){\textmd{$E$}}}
 \put(100,20){\framebox(20,15){\textmd{$I$}}}
 \put(20,27.5){\vector(1,0){30}}
 \put(70,27.5){\vector(1,0){30}}
 \put(120,27.5){\vector(1,0){20}}
 \put(110,35){\line(0,1){15}}
 \put(10,50){\vector(0,-1){15}}
 \put(10,50){\line(1,0){100}}
 \put(128,28){$d$}
 \put(24,30){$\alpha SE$}
 \put(24,18){$\beta SI$}
 \put(79,30){$\mu$}
 \put(60,53){$\lambda$}
\end{picture}
\centering \caption{Schematic flow diagram of transmission  for
SEIS model.}\label{myfigure04}
\end{minipage}

\par
Here the $\delta$ represents the maximum number individual of in
neighborhood for site $k$, then the number of sites in
$\delta_{k}$ is $\delta= |\delta_{k}|-1$, where the $\delta_{k}$
is the k-th neighborhood. There is no variation in size among the
sites. The local transition rule governs the infection
transmission about any site $k$. Let the mode of transmission of
the infection within $\delta_{k}$ like as follows: (i) if the k-th
site is non-empty, then we can assume that a exposed or infected
k-th site transmits the pathogen to a susceptible individual
within $\delta_{k}$ and (ii) if the k-th site is not infected or
exposed, then it acquires the pathogen ( infection) from a exposed
or infected individual within $\delta_{k}$. In our model, let us
assume that  $\delta_{k}$ is symmetric.  We also assume that the
site $k$ (or k-th site) with it's center is like as it is shown in
\autoref{myfigure1} and \autoref{myfigure2}. There are the same
transmit rule for all $\delta_{k}$. When the $\delta_{k}$ is
asymmetric, a host at site $k$ is infected from a sets differing
from the set to which the pathogen can be transmitted. Therefore
it will be difficult to discuss the mode of transmission in term
of mathematical equations. So for the sake of simplicity, we have
assumed that $\delta_{k}$ is symmetric. Now we define the  local
function $f$ as follows
\begin{equation}\label{EG01}
f:\longrightarrow z_{k}(t+1)=f(z_{k}(t),z_{\delta_{k}}(t)).
\end{equation}
Note that the spatial heterogeneity dispersion changes while the
local density and individual interaction neighborhoods are
changed. The discrete random variable $n_k$ counts the host in
$\delta_{k}$, where $0\leq n_k\leq\delta$. Then the mean number of
hosts per neighborhood is given by
\begin{equation}\label{EG02}
 E\big[n_{k}(t)\big]=\frac{N(t)\delta}{J}=H(t)\delta.
\end{equation}
\par

 The  $n_{k}$ hosts on $\delta_{k}$ include $s$
susceptibles, $e$ exposed and $i$ infectives; $n_k=s+e+i$. Here
$\alpha$ is the probability, per discrete time interval, that an
exposed on $\delta_{k}$ transmits the disease to the susceptible
at site $k$. $\beta$ is the probability, per discrete time
interval, that an infective on $\delta_{k}$ transmits the disease
to susceptible at site $k$. $\mu$ is the probability, per time
interval, that an exposed individuals convert to a infected one.
Since the infectives act independently, so the susceptible
acquires the pathogen in  a single time interval with conditional
probability $1-(1-\alpha)^{e}(1-\beta)^{i}$. The unconditional
probability of infecting the focal susceptible in a discrete time
interval is $\varepsilon_{k}$:
\begin{widetext}
$$\varepsilon_{k}=\sum^{\delta}_{n=1}\sum^{n}_{e=1}\sum^{n}_{i}\Bigl[1-(1-\alpha)^{e}
(1-\beta)^{i}\Bigr]\times Pr\big[e_{k}=e,i_{k}=i|n_{k}=n\big]\cdot
Pr\big[n_{k}=n\big].$$
\end{widetext}

\section{The approximation of the global density}

Mean-field analysis have different forms, here we adopt a
`local-dispersal' approximation~\cite{Knu8}. Let us assume $S(t)$,
E(t), I(t) are the total susceptible, exposed and infected
individuals  at time $t$ respectively. Then we have
$N(t)=S(t)+E(t)+I(t)$. The  global density of the exposed and the
infected populations  at time $t$ are given by
\begin{equation}\label{EQ02}
x(t)=\frac{E(t)}{J},\qquad y(t)= \frac{I(t)}{J}.
\end{equation}
Then, the global frequency of the exposed and the infected
populations at time $t$ are given by
\begin{equation}\label{EQ03}
\frac{E(t)}{N(t)}=\frac{E(t)}{H(t)J}=\frac{E(t)}{J}\;
\frac{1}{H(t)}=\frac{x(t)}{H(t)},\end{equation}
\begin{equation}\label{EQ04}
\frac{I(t)}{N(t)}=\frac{I(t)}{H(t)J}=\frac{I(t)}{J}\;\frac{1}{H(t)}=
\frac{y(t)}{H(t)}.
\end{equation}
For given independence of sites, the probability that a host
occupies a given site on $\delta_{k}$ is equal to the global
density $H(t)$. The individual interaction act as an independent
event in
 $\delta_{k}$, so the $(e,i)$ is an independent two-dimension
variable, and is denoted by $(\xi,\eta)$, where $\xi+\eta \leq
n_{k}$. Form the transmit rule $f$, we know that number of exposed
and infected individuals on $\delta_{k}$ has a (three-parameter)
binomial distribution with parameters $\delta,\; x(t),\; y(t)$. In
the following, the $\lambda$ is the probability, per time
interval, that an infective  recovers from infection and again
become an susceptible one. In the neighborhood $\delta_{k}$, the
exposed and infective at time $t+1$ and $t$ are independent.
Hence, we can get the global mean density of exposed and infective
at time $t+1$ are given as follows, where $x(t+1)$ represents the
exposed individuals and $y(t+1)$ represents the infected
individuals,
\begin{widetext}
\begin{eqnarray}\label{EQ05}
x(t+1)&=&F_{1}\Big(x(t),y(t)\Big)=x(t)(1-\mu)+\Big(H(t)-x(t)-y(t)\Big)\cdot\sum^{\delta}_{\xi=0}\sum^{\delta}_{\eta=0} \nonumber \\
&
&\cdot\tbinom{\delta}{\xi}\tbinom{\delta-\xi}{\eta}x(t)^{\xi}y(t)^{\eta}\Big(1-x(t)-y(t)\Big)^{\delta-\xi-\eta}
\bigl[1-(1-\alpha)^{\xi}\cdot(1-\beta)^{\eta}\bigr] \nonumber\\
&=&x(t)(1-\mu)+\Big(H(t)-x(t)-y(t)\Big)\cdot\Bigl[\sum^{\delta}_{\xi=0}\sum^{\delta}_{\eta=0}\tbinom{\delta}{\xi}\tbinom{\delta-\xi}{\eta}
x(y)^{\xi}y(t)^{\eta}\Big(1-x(t)-\nonumber  \\
& &y(t)\Big)^{\delta-\xi-\eta}
-\sum^{\delta}_{\xi=0}\sum^{\delta}_{\eta=0}\tbinom{\delta}{\xi}\tbinom{\delta-\xi}{\eta}x(y)^{\xi}y(t)^{\eta}\Big(1-x(t)-y(t)\Big)
^{\delta-\xi-\eta}(1-\alpha)^{\xi}(1-\beta)^{\eta}\Bigr] \nonumber \\
&=&x(t)(1-\mu)+\Big(H(t)-x(t)-y(t)\Big)\cdot\Bigl[1-\Big(1-\alpha
x(t)-\beta y(t)\Big)^{\delta}\Bigr].
\end{eqnarray}
\end{widetext}
and
\begin{eqnarray}\label{EQ06}
y(t+1)&=&F_{2}\Big(x(t),y(t)\Big)=y(t)(1-\lambda-d)+\mu x(t).
\end{eqnarray}\par

From Eq.~\eqref{EQ05} and Eq.~\eqref{EQ06} we can obtain that the
global density of exposed and infective are given as follows:
\begin{equation}\label{EQ07}
\begin{cases} x(t+1)=x(t)(1-\mu)+\Big(H(t)-x(t)-y(t)\Big)\\
\hspace{1.5cm}\Big[1-\Big(1-\alpha x(t)-\beta y(t)\Big)^{\delta}\Big], \\
y(t+1)=y(t)(1-\lambda-d)+\mu x(t).
\end{cases}
\end{equation} This is a two-dimension nonlinear system. Since
$D\ll N(t)$ for not too long time, so we can omit the change of
total individual (see~\cite{ODk}) and let $H(t)=H=\rm constant$,
$x(t+1)=x(t)=x,$ and $y(t+1)=y(t)=y$. Where the symbol $D$
represents the number of death up to time $t$. The equilibrium
points of the system (\ref{EQ07}) satisfy the equations
\begin{equation}\label{EQ08}
\begin{cases} x=x\cdot(1-\mu)+(H-x-y)\cdot\big[1-(1-\alpha x-\beta y)^{\delta}\big],\\
y=y\cdot(1-\gamma)+\mu x(t).
\end{cases}
\end{equation}
Now in the above system , we have $\gamma=d+\lambda$. Since
$\delta>0$, $0\leq x(t)<H$,\;$0\leq y(t)<H$, so solving these
above equations we get a zeros-fixed point $(x_{0},y_{0})$:
$$
\begin{cases}x_{0}=0,\\
y_{0}=0.
\end{cases}
$$
Now we consider the Jacobian matrix of the system~(\ref{EQ07}) at
$(0,0)$. From the equation ~(\ref{EQ05}) and the equation
~(\ref{EQ06}) we obtain
\begin{widetext}
\begin{eqnarray}\label{EQ10}
  \frac{\partial F_{1}}{\partial x(t)} &=& (1-\alpha x-\beta y)^{\delta}+\alpha\delta(H-x-y)(1-\alpha
x-\beta y)^{\delta-1}-\mu, \nonumber\\
 \frac{\partial F_{1}}{\partial y(t)} &=&(1-\alpha x-\beta y)^{\delta}+\beta\delta(H-x-y)(1-\alpha
x-\beta y)^{\delta-1}-1,\nonumber\\
 \frac{\partial F_{2}}{\partial x(t)} &=& \mu,\nonumber \\
 \frac{\partial F_{2}}{\partial y(t)} &=& 1-\gamma .\nonumber
\end{eqnarray}
\end{widetext} Therefore the Jacobian Matrix $\mathbf{A}$ can be
expressed at the disease free equilibrium $(0,0)$ as follows
\begin{equation*}\label{abc}
 \textbf{A}=\left(%
\begin{array}{cc}
  1+H\alpha\delta-\mu & H\beta\delta \\
 \mu & 1-\gamma \\
\end{array}%
\right).
\end{equation*} Then the trace and determinant
of the matrix $\textbf{A}$ are given by
\begin{equation}\label{EQ11}
 \texttt{tr}\textbf{A}=2-\mu-\gamma+H\delta\alpha
\end{equation} and
\begin{eqnarray}\label{EQ12}
  \det\textbf{A} &=& (1-\mu+H\delta\alpha)(1-\gamma)-H\delta\beta\mu \nonumber \\
  &=&
  1-\mu+H\delta\alpha-\gamma+\mu\gamma-H\delta(\alpha\gamma+\beta\mu).
\end{eqnarray}

 Now let us denote
$\mathcal{R}_c=\frac{H\delta(\alpha\gamma+\beta\mu)}{\gamma\mu}$,
as the persister threshold, then the following theorem holds true.

\textbf{Theorem 1.} \emph{If $\mathcal{R}_c<1$, then the system
(\ref{EQ07}) is locally asymptotically stable around the disease
free equilibrium $(0,0)$ and unsatble if $\mathcal{R}_c>1$.}
\begin{proof}
Let us linearize the system~(\ref{EQ07}) at $(0,0)$, we get
\begin{equation}\label{EQ14}
\begin{cases} x(t+1)=(1+H\alpha\delta-\mu)x(t)+H\beta\delta y(t), \\
y(t+1)=\mu x(t)+(1-\gamma)y(t).
\end{cases}
\end{equation}

 Let us consider a positive Lyapunov function $V =
(\frac{\alpha}{\mu}+\frac{\beta}{\gamma})x+\frac{\beta}{\gamma}y$,
then the time derivative of this function along the solution of
the system \eqref{EQ14}, we get
\begin{eqnarray*}
\Delta V &=& V(t+1)-V(t) \\
&=&
(\frac{\alpha}{\mu}+\frac{\beta}{\gamma})\big[x(t+1)-x(t)\big]+\frac{\beta}{\gamma}\big[y(t+1)-y(t)\big]\\
&=& -(\frac{\alpha}{\mu}+\frac{\beta}{\gamma})\mu
x+(\frac{\alpha}{\mu}+\frac{\beta}{\gamma})H\delta(\alpha x+\beta
y)-\beta y+\frac{\beta}{\gamma}\mu x\\
&=& -\alpha x-\beta
y+(\frac{\alpha}{\mu}+\frac{\beta}{\gamma})H\delta(\alpha x+\beta
y)\\
&=& -(\alpha x+\beta y)(1-\mathcal{R}_c)< 0.
\end{eqnarray*}

 Again  note that $\Delta V=0$ at the origin $(0, 0)$.
Thus, according to  Lyapunov LaSalle Theorem~\cite{Knu12}, the
origin of system \eqref{EQ07} is locally asymptotically stable.

 Now if $\mathcal{R}_c >1$, one of the Jure conditions
~\cite{Knu13} is violated as
\begin{equation}\label{}
  1-\texttt{tr}\textbf{A}+\texttt{det}\textbf{A}=\gamma\mu-H\delta(\alpha\gamma+\beta\mu)>0,
\end{equation}
which implies that the matrix  $ \textbf{A}$ has a
 pair of complex-conjugate eigenvalues lying outside the unit
circle, which again implies that the equilibrium $(0,0)$ of the
system  (\ref{EQ07}) is unstable.
\end{proof}

 The above theorem gives us  the condition for which the
pathogen will extinct and when it will persist. The system
\eqref{EQ07} has an infection-free equilibrium in which the
component of infectives is zero and exposed is zero. After
analyzing the local stability of the system \eqref{EQ07} around
$(0,0)$ we get a threshold value  which determine when the disease
dies out and when persist in the system. These threshold
conditions are characterized by the critical threshold
$\mathcal{R}_c $, such that $(0,0)$ is locally asymptotically
stable if $\mathcal{R}_c<1 $, and unstable if $\mathcal{R}_c
>1$. The numerical simulation shows this fact for different values of
$\mathcal{R}_c$.\\
\setcounter{figure}{3}
\setcaptionwidth{8cm}

\begin{minipage}[b]{0.48\textwidth}
\subfigure[$\mathcal{R}_{c}=0.4495$]{%
\label{fig0:a}
\includegraphics[width=0.45\textwidth]{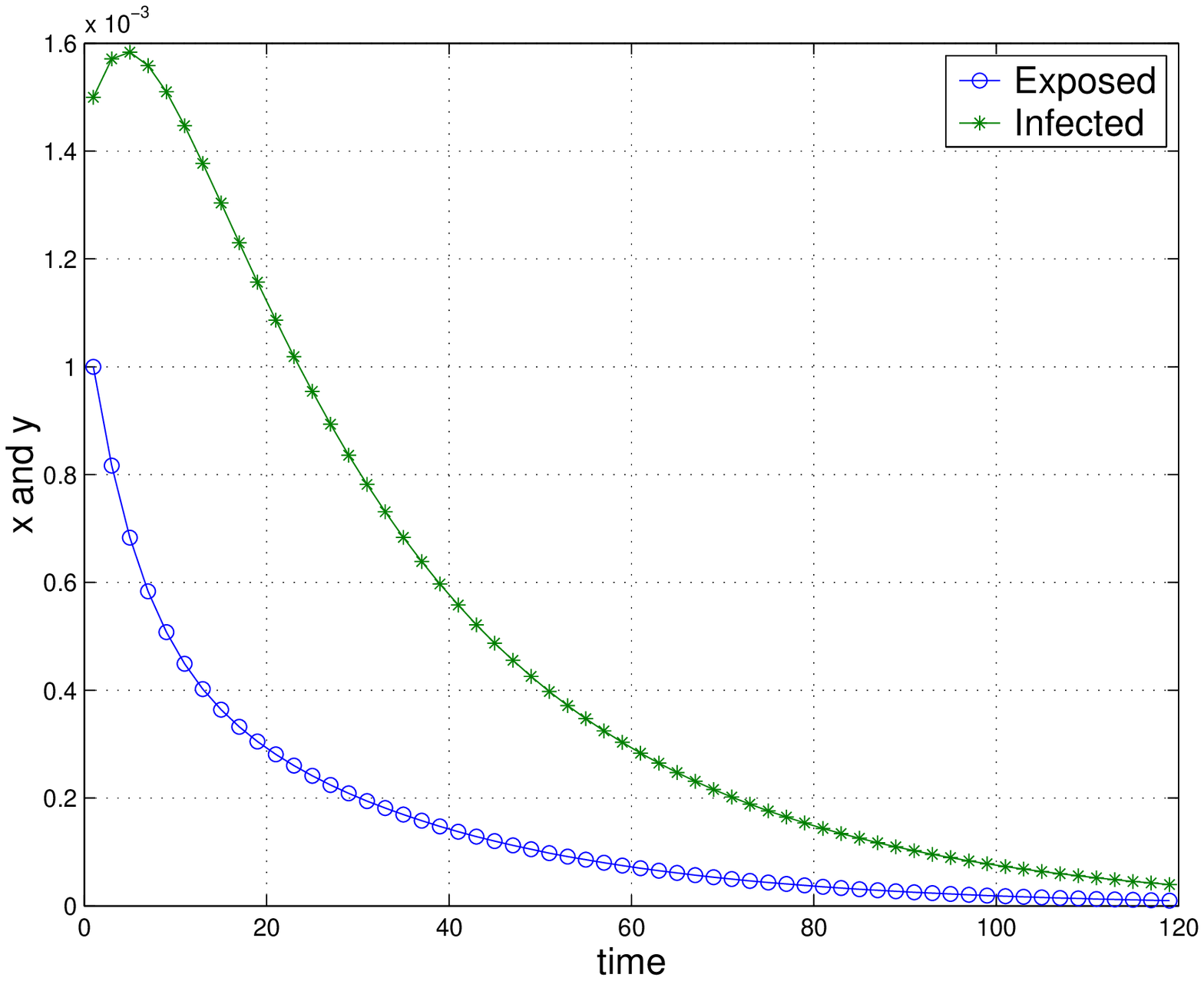}}
\subfigure[$\mathcal{R}_{c}=1$]{
\label{fig0:a}
\includegraphics[width=0.45\textwidth]{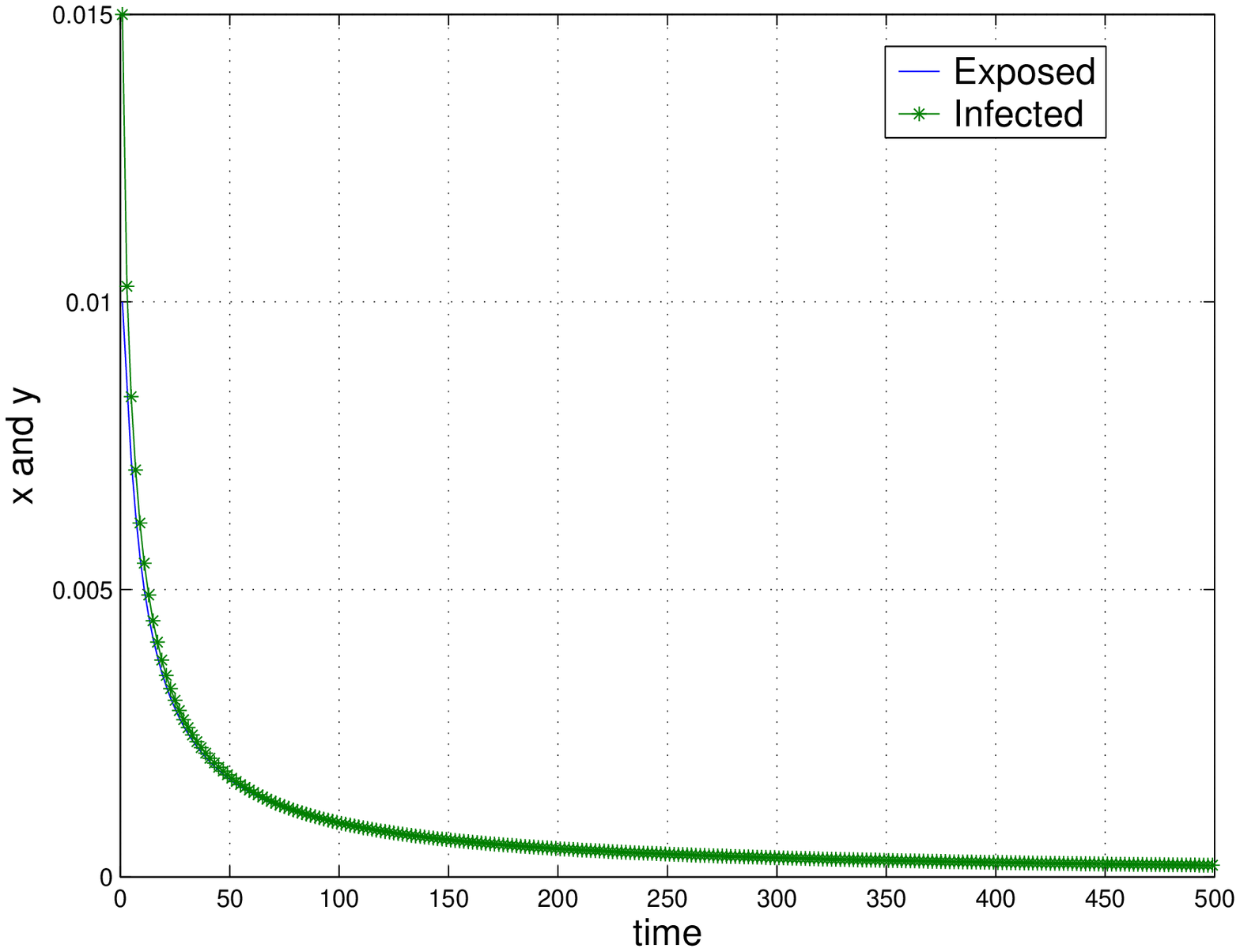}}\\
\subfigure[$\mathcal{R}_{c}=7.9653$]{\label{fig0:a}
\includegraphics[width=0.45\textwidth]{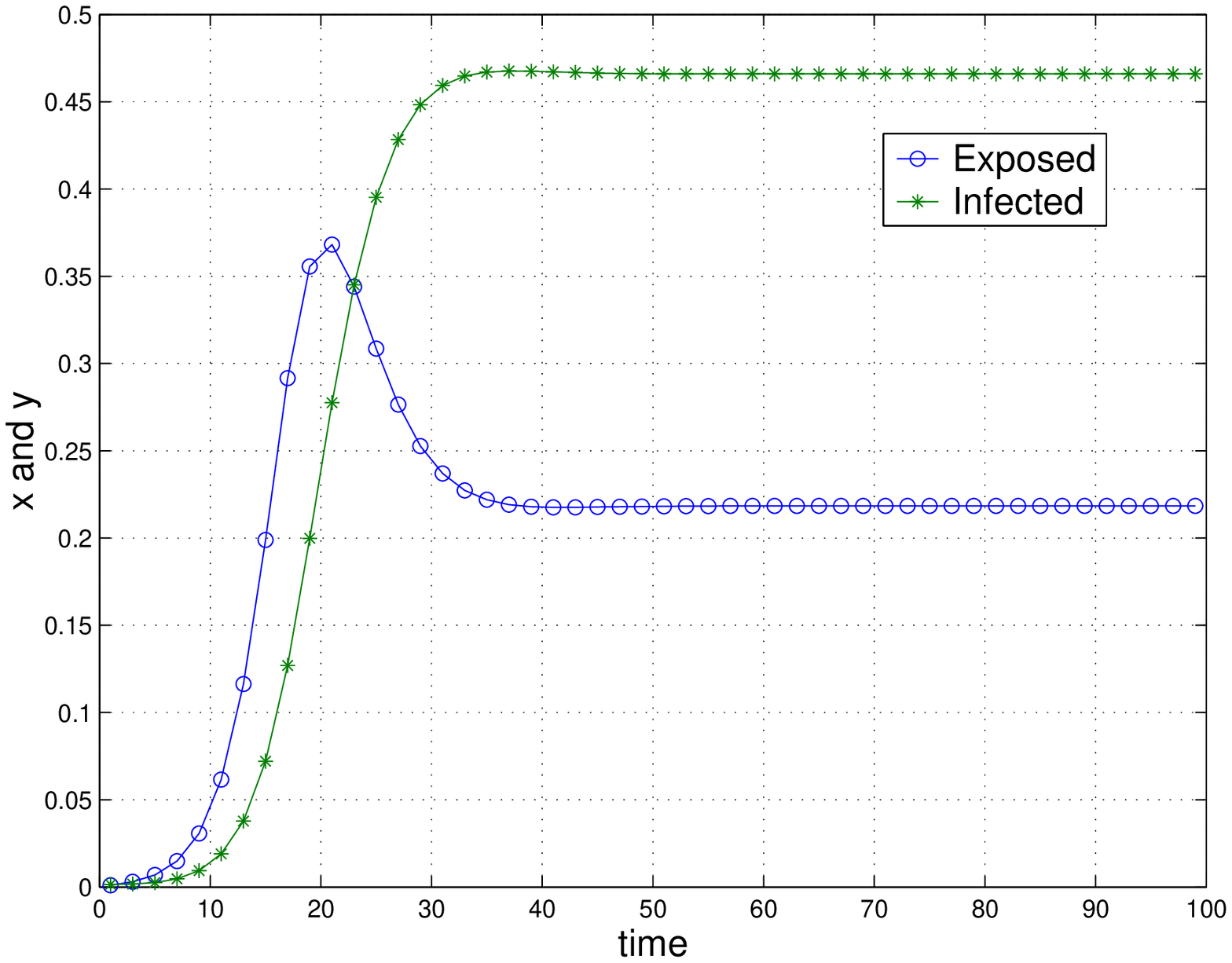}}
\subfigure[$\mathcal{R}_{c}=1.9092$]{\label{fig0:a}
\includegraphics[width=0.45\textwidth]{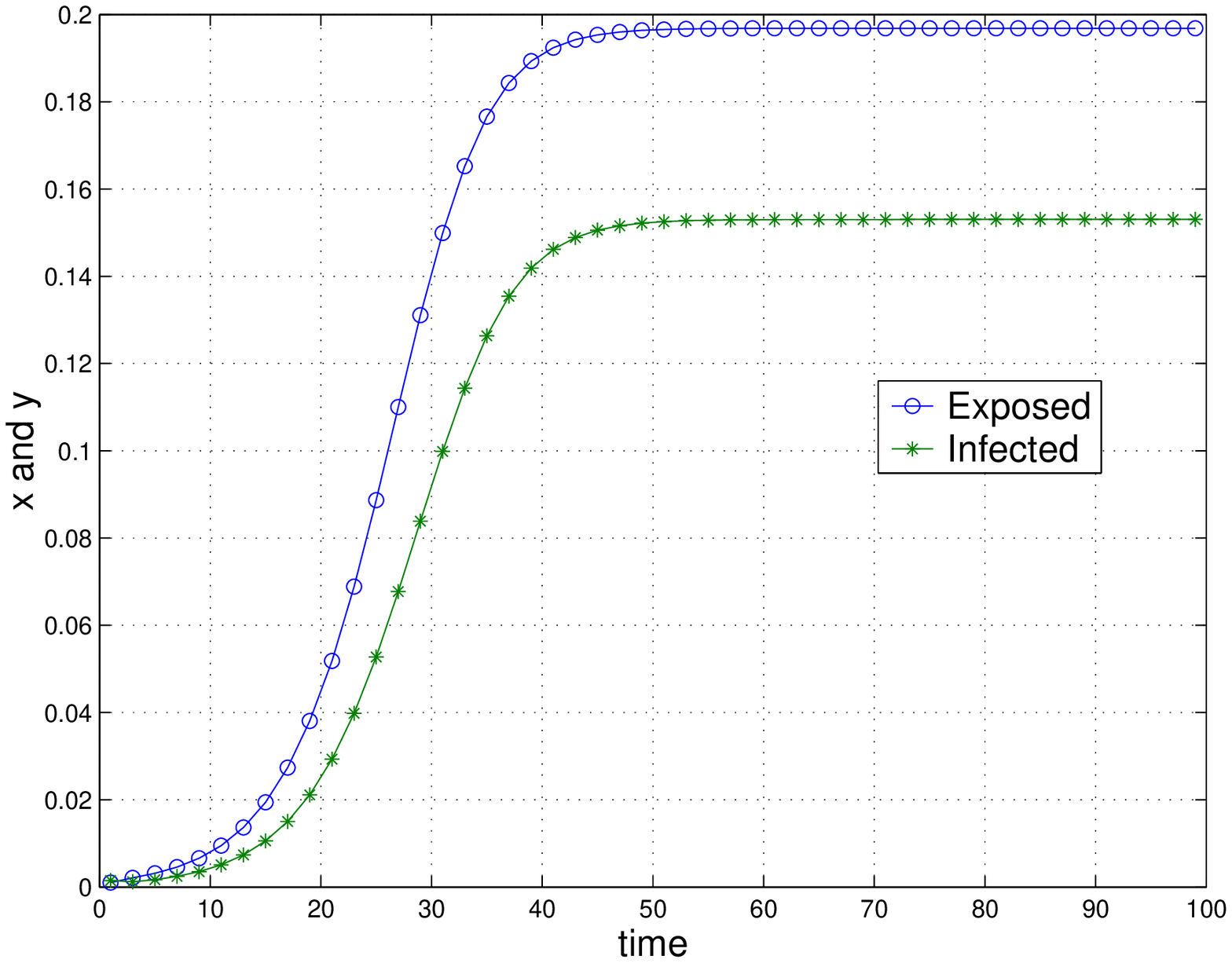}}\\
\subfigure[$\mathcal{R}_{c}=1.0011$]{\label{fig0:a}
\includegraphics[width=0.45\textwidth]{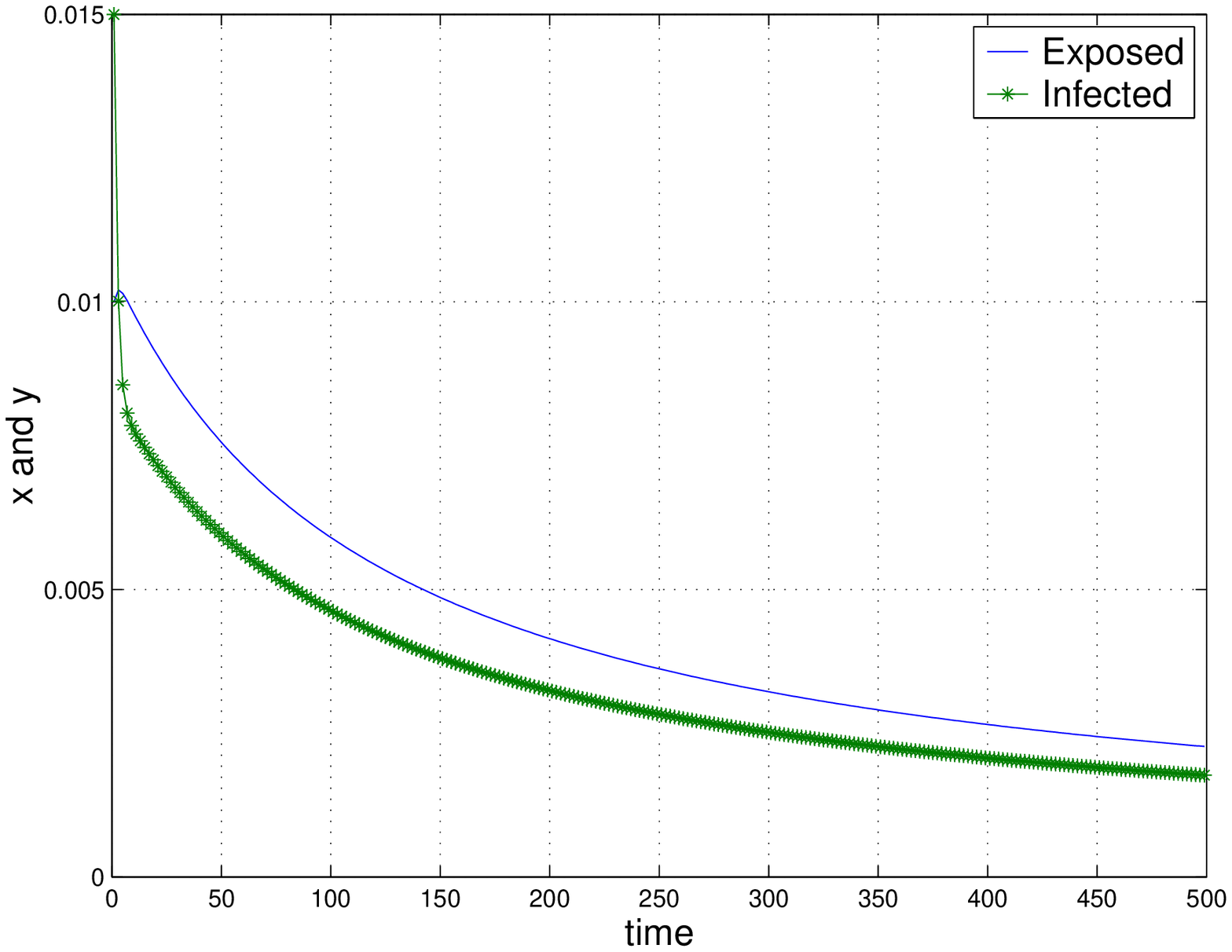}}
\caption{\small{ For each figures we assume the following initial
condition: $x(0)=0.001$, $y(0)=0.0015$. (a) Figure a: Figure a
have drawn with the parameters values: $\alpha=0.002,
\beta=0.004,\mu=0.15,\delta=8,H=0.8,d=0.0003$ and $\lambda=0.07$.
(b) For figure b: $\alpha=\beta=\mu=0.64, d=0.003$, and
$\lambda=0.637$. (c) For figure c: with the same parameters value
that in figure (a) except $\alpha=0.08, \beta=0.05$. (d) For
figure d: $\mathcal{R}_{c}=1.9092, \mu=0.35, \alpha=0.5,
\beta=0.07$ and $\lambda=0.45$, the value of other parameters are
remaining same as in figure (a). (e)  For figure e:
$\mathcal{R}_c=1.0011, \mu=0.35, \alpha=0.05, \beta=0.0061105,
x(0)=0.01$ and $y(0)=0.015$ and $\lambda=0.45$, the value of other
parameters are remaining same with figure (a).}}\label{myfigure0}
\vspace{0.2cm}
\end{minipage}


 Now we are in position to show behaviour of the system
(\ref{EQ07}) around the positive interior equilibrium.

\textbf{Theorem 2.} \emph{If
$\mathcal{R}_c=\frac{H\delta(\alpha\gamma+\beta\mu)}{\gamma\mu}>1$,
the system (\ref{EQ07}) possess a  positive interior equilibrium
$(x^{*},y^{*})$.}
\begin{proof}

The system of (\ref{EQ08}) can be expressed as
\begin{equation}\label{EQ13}
\begin{cases} \mu x=(H-x-y)\big[1-(1-\alpha x-\beta y)^{\delta}\big],\\
y=hx.
\end{cases}
\end{equation}
Where $h=\mu/\gamma$. Again the equation~(\ref{EQ13}) can be
written as
$$\mu x=(H-x-hx)\big[1-(1-\alpha x-\beta hx)^{\delta}\big].$$ Now let
us assume $g(x)=(H-x-hx)\big[1-(1-\alpha x-\beta
hx)^{\delta}\big]-\mu x$.
 Since $0<\alpha<1, 0<\beta<1, 0\leq x\leq 1$ and $0\leq y\leq 1$,
 so $g(0)=0$ and $dg(x)/dx$ is given by
 \begin{eqnarray*}
    g^{'}(x)&=& -(1+h)\big[1-(1-\alpha x-\beta hx)^{\delta}
 \big]+ \\
   & & (H-x-hx)\delta (\alpha+\beta h)(1-\alpha x-\beta
 hx)^{\delta-1}-\mu.
 \end{eqnarray*}
Therefore
\begin{eqnarray*}
 g^{'}(0) &=&H\delta(\alpha+\beta h)-\mu\\
  &=& H\delta(\alpha+\frac{\mu\beta}{\gamma})-\mu \\
  &=&
  \frac{1}{\gamma}\big[H\delta(\alpha\gamma+\beta\mu)-\mu\gamma\big]>0.
\end{eqnarray*}
So, the function $g(x)$ is increasing at $(0,0)$. Therefore if we
take $0\leq \tau=\frac{H}{1+h}<1$ , then we get
$$g(\tau)=-\frac{H\mu}{1+h}<0.$$

Thus, there should be positive interior equilibrium of the system
in the open interval (0, $\frac{H\mu}{1+h}$). Hence the proof.
\end{proof}

 Now we are interested to find out the positive interior
 equilibrium of the system in open interval (0,
 $\frac{H\mu}{1+h}$). Now for small values of $\alpha$ and $\beta$, taking Taylor
expansion and keeping only the fist two terms, we get
\begin{equation}\label{EQ23}
 (1-\alpha x-\beta y)^{\delta}=1-\delta(\alpha x+\beta y)+\cdots.
\end{equation}

So from the systems  (\ref{EQ08}), we obtain

\begin{equation}\label{EQ24}
\begin{cases} x=x(1-\mu)+\delta(H-x-y)(\alpha x+\beta y),\\
y=y(1-\gamma)+\mu x.
\end{cases}
\end{equation}
Therefore when $\mathcal{R}_c>1$, we can derive the positive
interior equilibrium of the system \eqref{EQ24} as follows

\begin{equation}\label{EQ25}
 \begin{cases}
    x^{*}=\frac{H\gamma}{\gamma+\mu}-\frac{\gamma^{2}\mu}{\delta(\gamma+\mu)(\alpha\gamma+\beta\mu)},\\
    y^{*}=\frac{H\mu}{\gamma+\mu}-\frac{\mu^{2} \gamma}{\delta(\gamma+\mu)(\alpha\gamma+\beta\mu)}.\\
  \end{cases}
\end{equation}
Together with the equation~(\ref{EQ03}) and the
equation~(\ref{EQ04}), we get
\begin{equation}\label{EQ26}
\begin{cases}
    \frac{x^{*}}{H}=\frac{\gamma}{\gamma+\mu}-\frac{\gamma^2\mu}{H\delta(\gamma+\mu)(\alpha\gamma+\beta\mu)},\\
    \frac{y^{*}}{H}=\frac{\mu}{\gamma+\mu}-\frac{\gamma\mu^2}{H\delta(\gamma+\mu)(\alpha\gamma+\beta\mu)}.
  \end{cases}
\end{equation}
From the equation \eqref{EQ26} we can conclude that when
$\mathcal{R}_c>1$, then the global frequency of exposed and
infective $(\frac{x^*}{H},\frac{y^*}{H})$ increases with
neighborhood size $\delta$ and with the pathogen transmission
probability $\alpha$ and $\beta$. Figure 6 shows this fact. Again
from \autoref{myfigure5}, we can conclude that infections
decreases with the increasing of the recovery probability
$\lambda$ ( see Figure 7(a))  but the size of the exposed
individuals will increase with the increasing of $\lambda$ ( see
Figure 7(b)) . This observation supports the observation of
Ellner~\cite{Knu16}.
\par

 Now Since
$0<\alpha<1, 0<\beta<1, 0\leq x\leq 1$ and $0\leq y\leq 1$, so
$0\leq(\alpha x+\beta y)<1$. Now  if the size of the interaction
neighborhood and the environment increases, that is,
$\delta\rightarrow J, J\rightarrow \infty$, then we get
\begin{equation}\label{EQ16}
 1-(1-\alpha x-\beta y)^{\delta}\rightarrow 1,
\end{equation}
as a consequence the equation (\ref{EQ08}) reduces to
\begin{equation}\label{EQ17}
\begin{cases} x=x(1-\mu)+(H-x-y),\\
y=y(1-\gamma)+\mu x.
\end{cases}
\end{equation}
Now solving the above equation and denoting the positive interior
equilibrium as $(x^+, y^+)$ we obtain
\begin{equation}\label{EQ18}
\begin{cases}
    x^{+}=\frac{H\gamma}{\mu\gamma+\gamma+\mu},\\
    y^{+}=\frac{H\mu}{\mu\gamma+\gamma+\mu},
  \end{cases}
\end{equation}
\par
\noindent which can be written as follows
\begin{equation}\label{EQ21}
\begin{cases}
    \frac{x^+}{H}=\frac{\gamma}{\mu\gamma+\gamma+\mu},\\
    \frac{y^+}{H}=\frac{\mu}{\mu\gamma+\gamma+\mu}.
  \end{cases}
\end{equation}
\par

Therefore the frequency of exposed and infected hosts are
approximated to $([1+\mu+\frac{\mu}{d+\lambda}]^{-1},
 [1+\gamma+\frac{d+\lambda}{\mu}]^{-1})$. Now the by comparison of the solution \eqref{EQ26}
 with the solution  \eqref{EQ21}, it is easy to show that they are approximately equal if $\delta\rightarrow \infty$ .
 Note that the solution  \eqref{EQ26} is better than the solution  \eqref{EQ21}, which shows the relation between exposed,  infected and susceptible
 individuals.

\textbf{Half von Neunann neighborhood:} Let us consider a special
case when $\delta=2$,  the neighborhood become one-dimensional
which is half of the standard von Neunann neighborhood. Here in
our case the neighborhood have only horizontal neighbors or
isotropic neighbors, and it have the qual probability for each
step. Hence we define the radius $r=1/2$, the equation of
\eqref{EQ08} can be written as
\begin{equation}\label{EQ22}
\begin{cases} x=x(1-\mu)+(H-x-y)\big[1-(1-\alpha x-\beta y)^{2}\big],\\
y=y(1-\gamma)+\mu x.
\end{cases}
\end{equation}
Solving the equation (\ref{EQ22}) we get
$$
  \begin{cases}
    x_{1}=\frac{\gamma((2+H\alpha)\gamma+(2+H\beta)\mu-\sqrt{A})}{2(\gamma+\mu)(\alpha\gamma+\beta\mu)},\\
    y_{1}=\frac{\mu((2+H\alpha)\gamma+(2+H\beta)\mu-\sqrt{A})}{2(\gamma+\mu)(\alpha\gamma+\beta\mu)}.\\
    \end{cases}
$$
$$
  \begin{cases}
    x_{2}=\frac{\gamma\big((2+H\alpha)\gamma+(2+H\beta)\mu+\sqrt{A}\big)}{2(\gamma+\mu)(\alpha\gamma+\beta\mu)},\\
    y_{2}=\frac{\mu\big((2+H\alpha)\gamma+(2+H\beta)\mu+\sqrt{A}\big)}{2(\gamma+\mu)(\alpha\gamma+\beta\mu)},\\
    \end{cases}
$$
where
\begin{eqnarray*}
  A &=& (H\beta-2)^2\mu^2+\gamma^2(4-4H\alpha+H^2\alpha^2+4\mu)+ \\
  & & 2\gamma\mu(H^2\alpha\beta-2H(\alpha+
\beta)+2(2+\mu)).
\end{eqnarray*}

\section{Numerical simulations}

\begin{figure}[h]
\subfigure[]{
\label{fig4:a}\includegraphics[width=4cm]{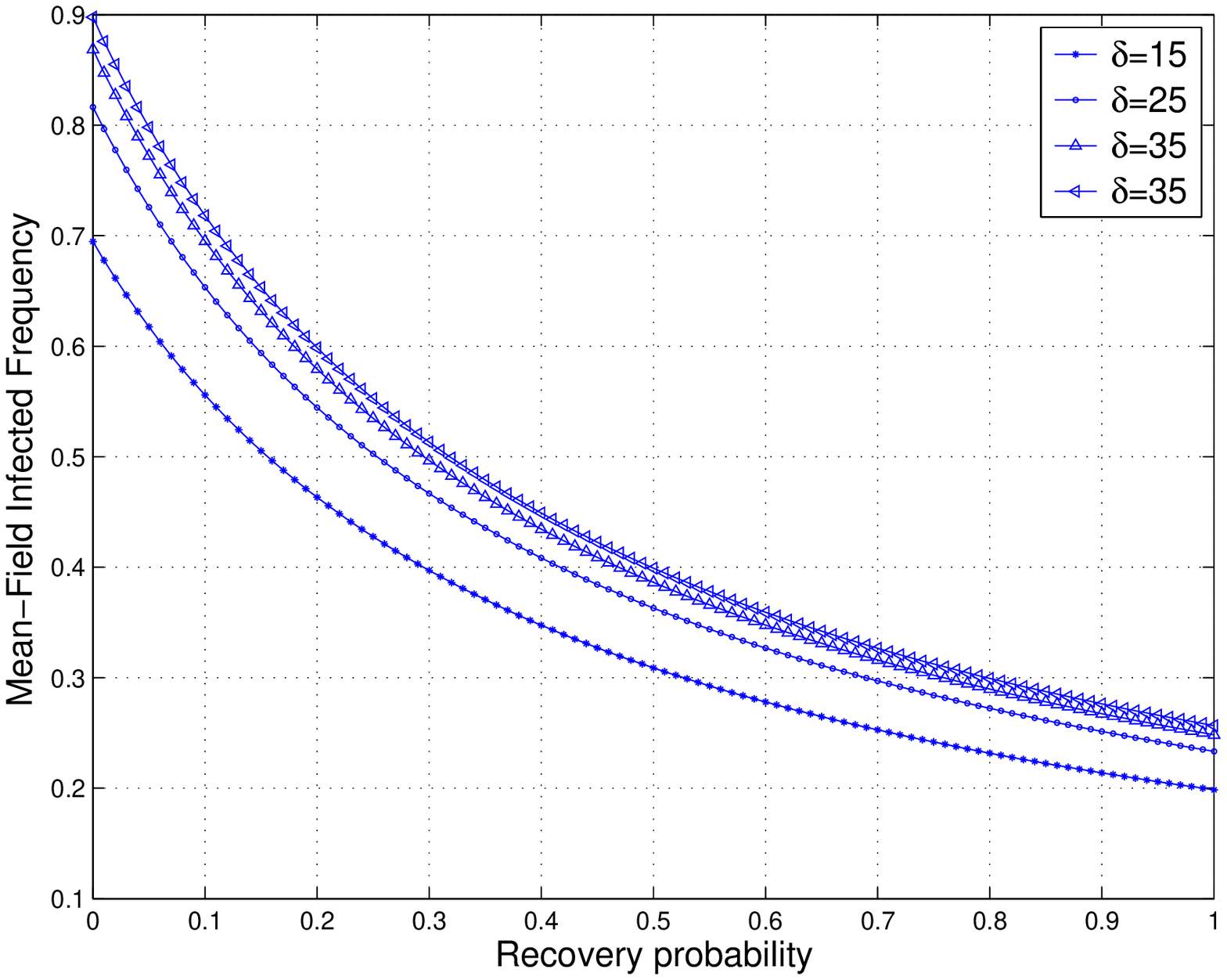}}
\hfill
\subfigure[]{\label{fig4:b}\includegraphics[width=4cm]{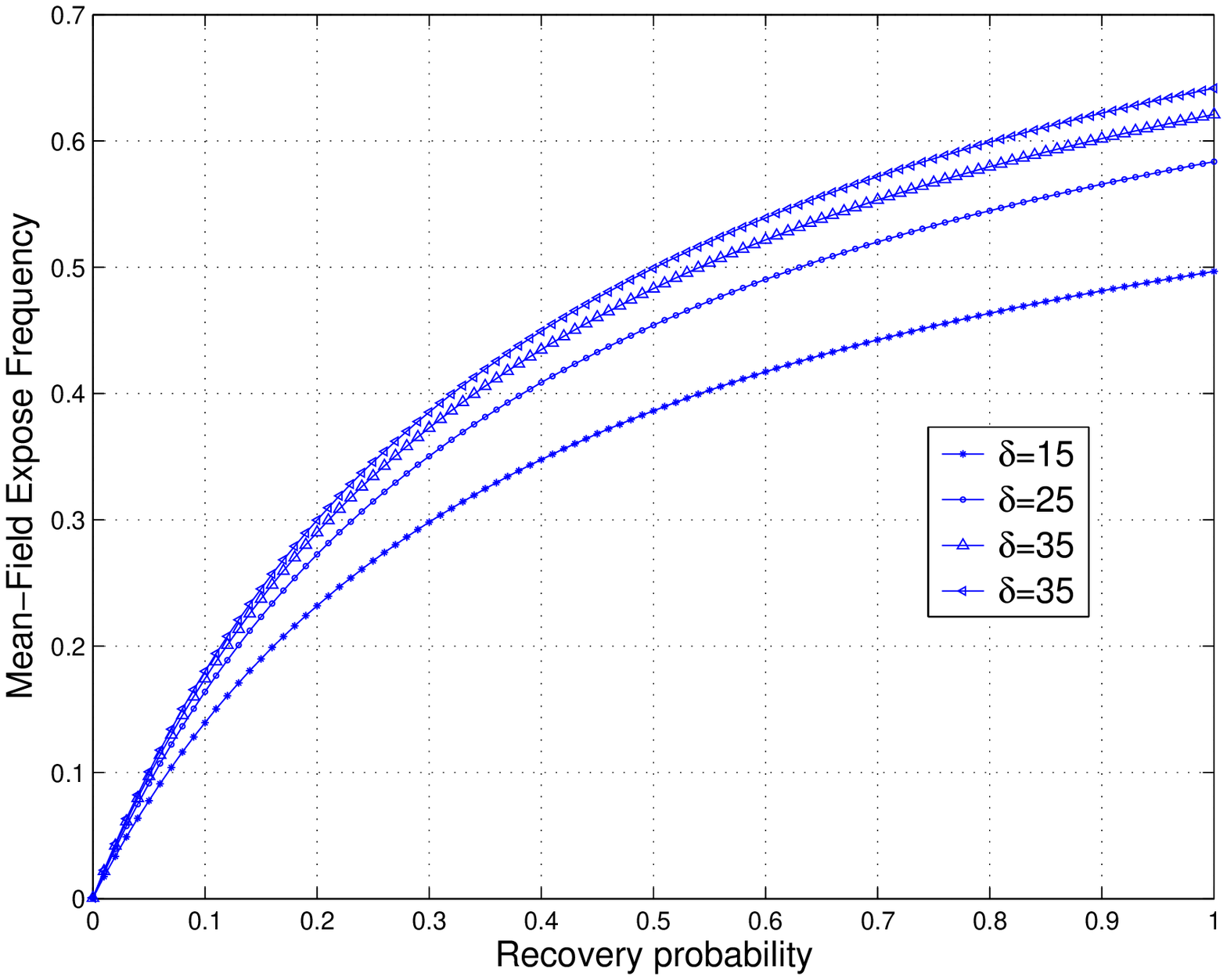}}
\caption{The effects of recovery probability.
 The above figures are drawn for different values of $\delta$ with
 $\;H=0.7$, $\mu=0.8$, $\alpha=1/6$, $d=0.0003$ and
$\beta=1/4$.}\label{myfigure5}
\end{figure}

In this section we have shown the numerical simulation of our CA
model with the help Matlab  programme. Here we have considered the
case for a common place like any public organization, school,
company and so on, hence we assume $J=10000$. We have also
simulated for a large value of $J$ but have the similar results
like in our other paper~\cite{Knu19}. The sites are updated
synchronously. \autoref{myfigure4} shows that  the global
frequency of exposed and infected individuals  increase with the
increasing of the size of the neighborhood provided
$\mathcal{R}_c> 1$.  The \autoref{myfigure5} shows that infections
decreases with the increasing of the recovery probability
$\lambda$ ( see Figure 7(a))  but the size of the exposed
individuals will increase with the increasing of $\lambda$ ( see
Figure 7(b)) . \autoref{myfigure6} and \autoref{myfigure7} are
showing the role of spatial heterogeneity  with $\delta=8$. The
Figure 8(a) and figure 8(c) are showing the disorder
dispersions of the hosts at initial time.\\
\begin{minipage}[b]{0.48\textwidth}
\subfigure[]{
\label{fig4:a}\includegraphics[width=0.45\textwidth]{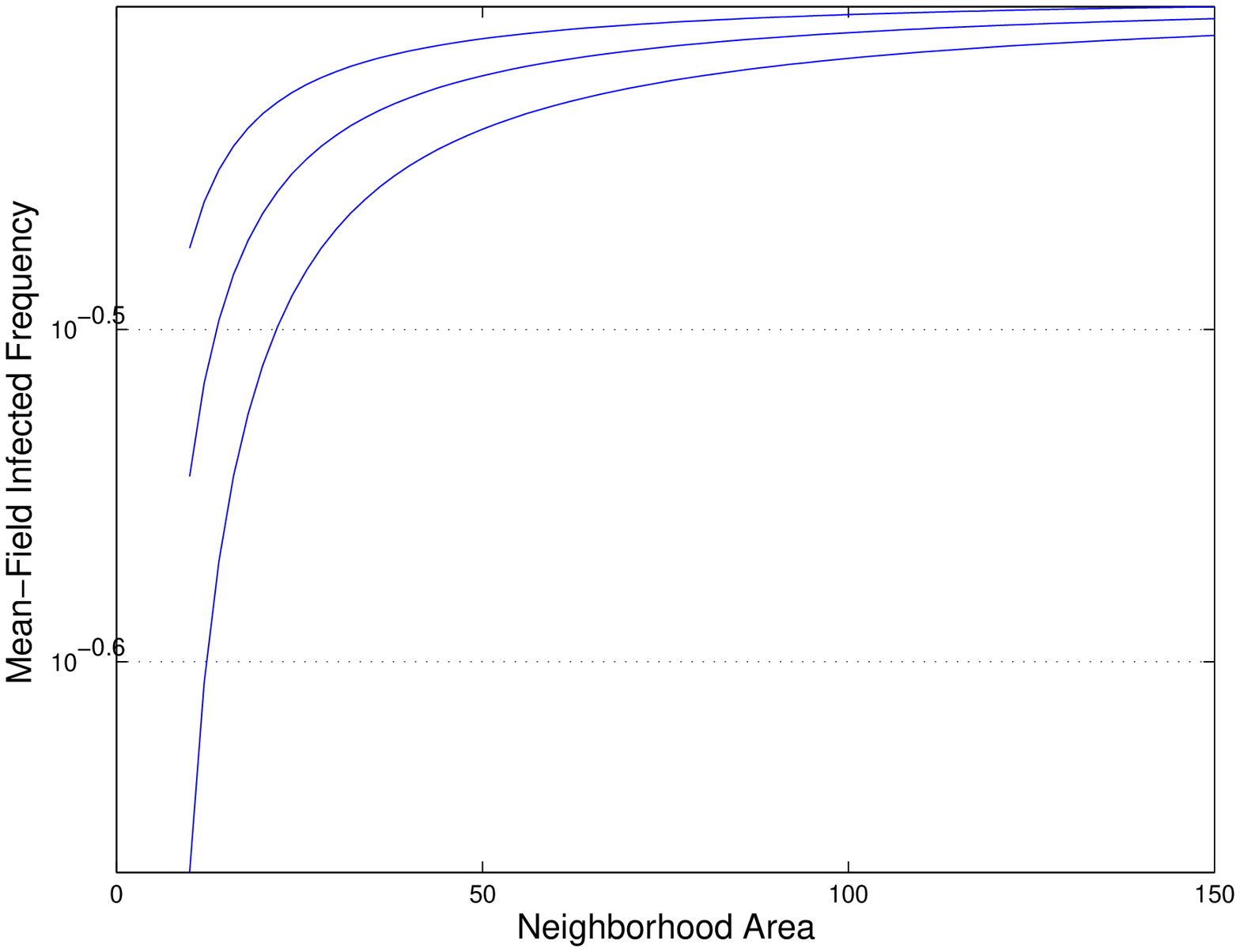}}
\hfill
\subfigure[]{\label{fig4:a}\includegraphics[width=0.45\textwidth]{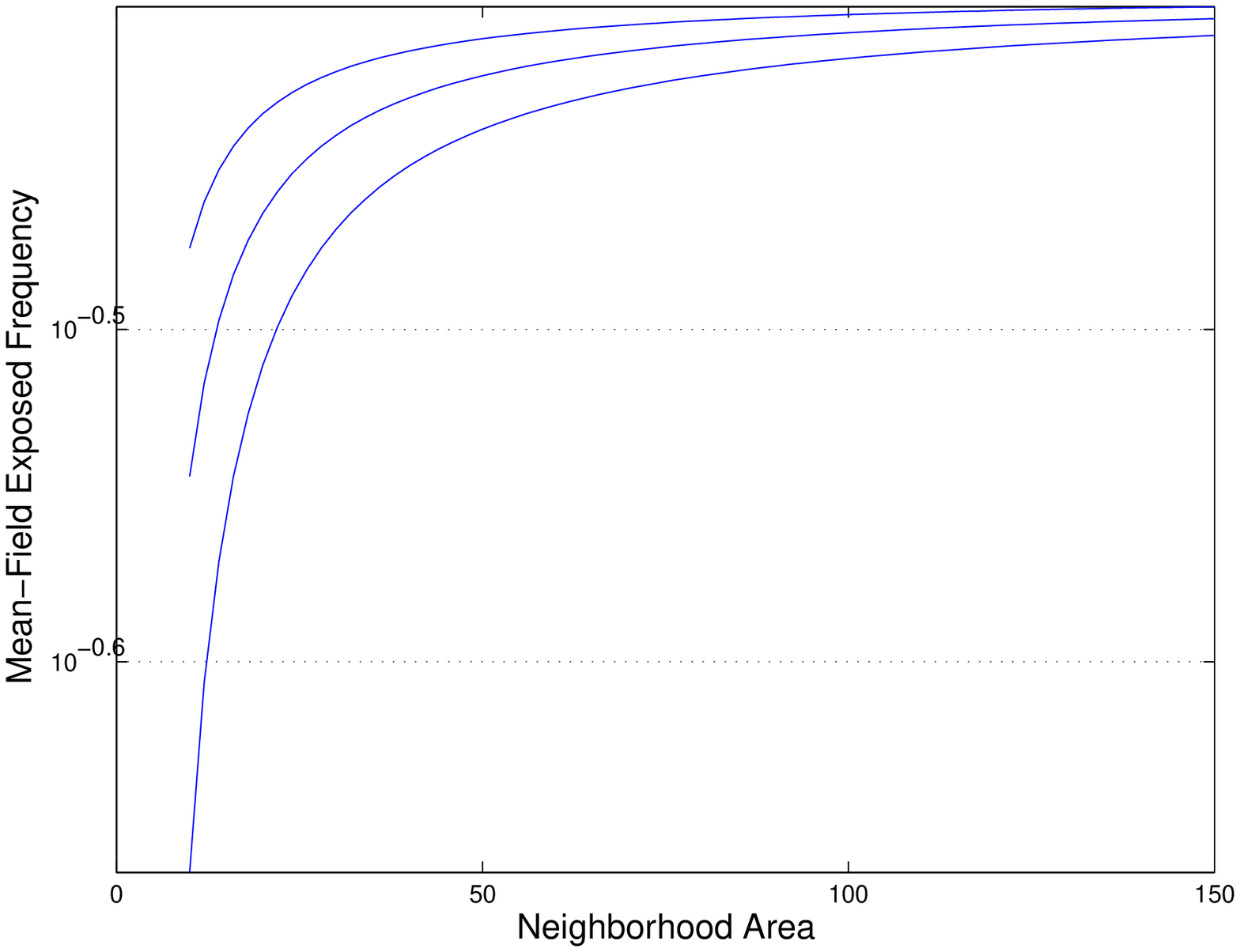}}
\caption{The frequency of exposed and infected individuals
 with a logarithmic scale at mean-field equilibrium. The parameters values are: H=0.7, $\lambda=0.6,
d=0.0003, \mu=0.4$ for each figure. In the above figures, the
upper line is drawn with $\alpha=\frac{1}{4}$ and
$\beta=\frac{1}{2}$; the middle line with $\alpha=\frac{1}{6}$ and
$\beta=\frac{1}{4}$; the lowermost line with $\alpha=\frac{1}{8}$
and $\beta=\frac{1}{8}$.}\label{myfigure4}
\end{minipage}
\begin{minipage}[b]{0.48\textwidth}
\subfigure[]{
\label{fig6:a}\includegraphics[width=0.45\textwidth]{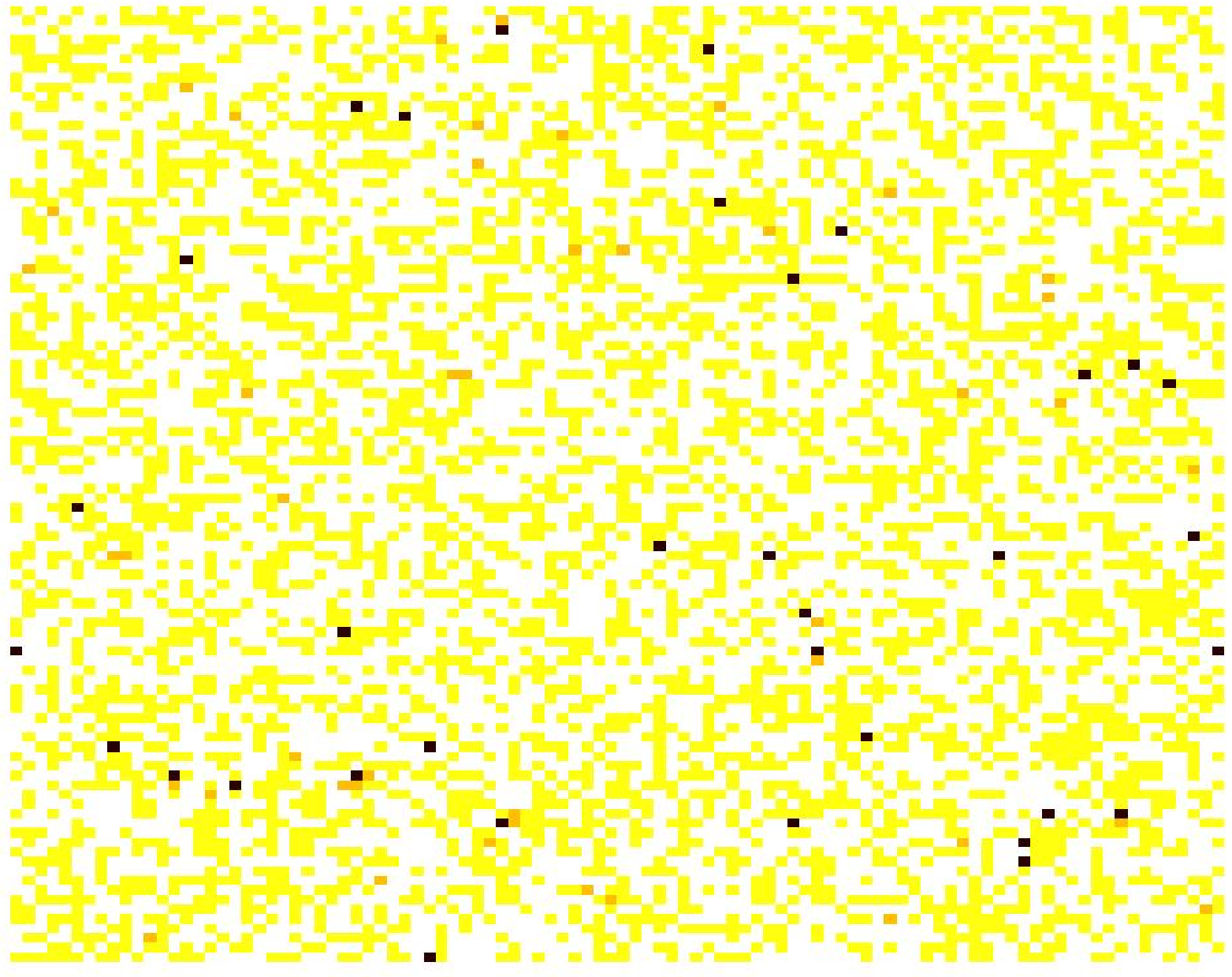}}
\subfigure[]{\label{fig6:a}\includegraphics[width=0.45\textwidth]{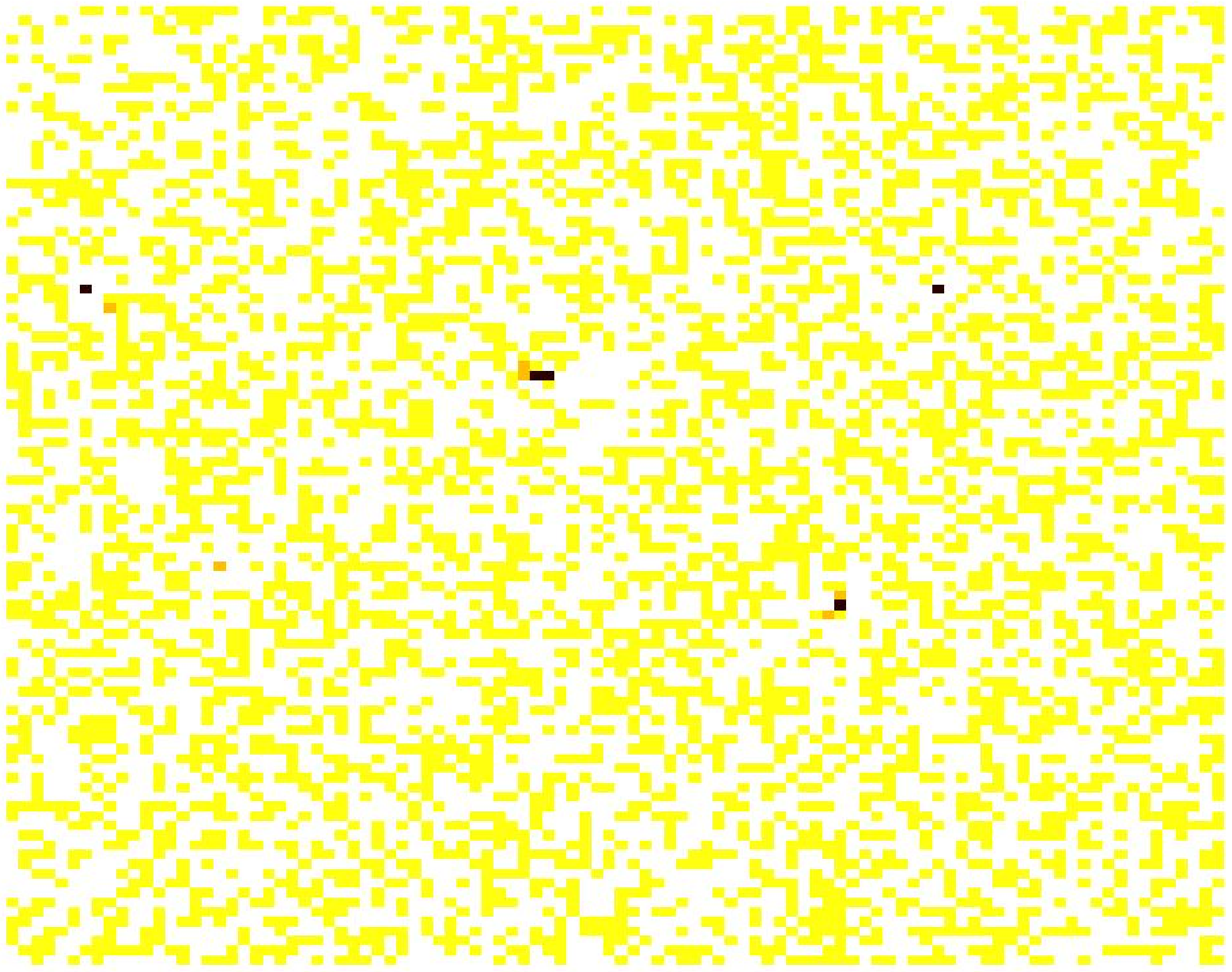}}\\
\subfigure[]{\label{fig6:a}\includegraphics[width=0.45\textwidth]{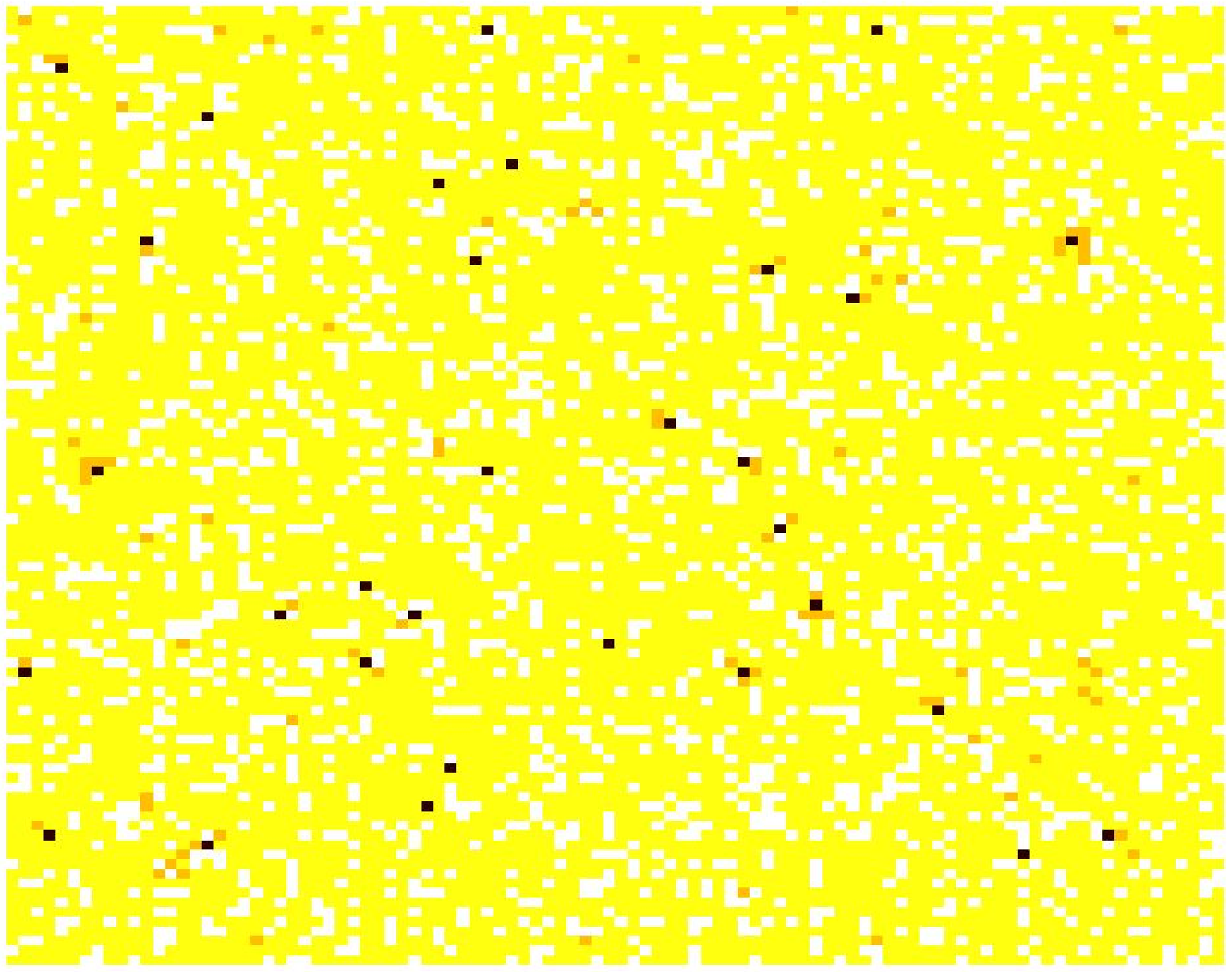}}
\subfigure[]{\label{fig6:a}\includegraphics[width=0.45\textwidth]{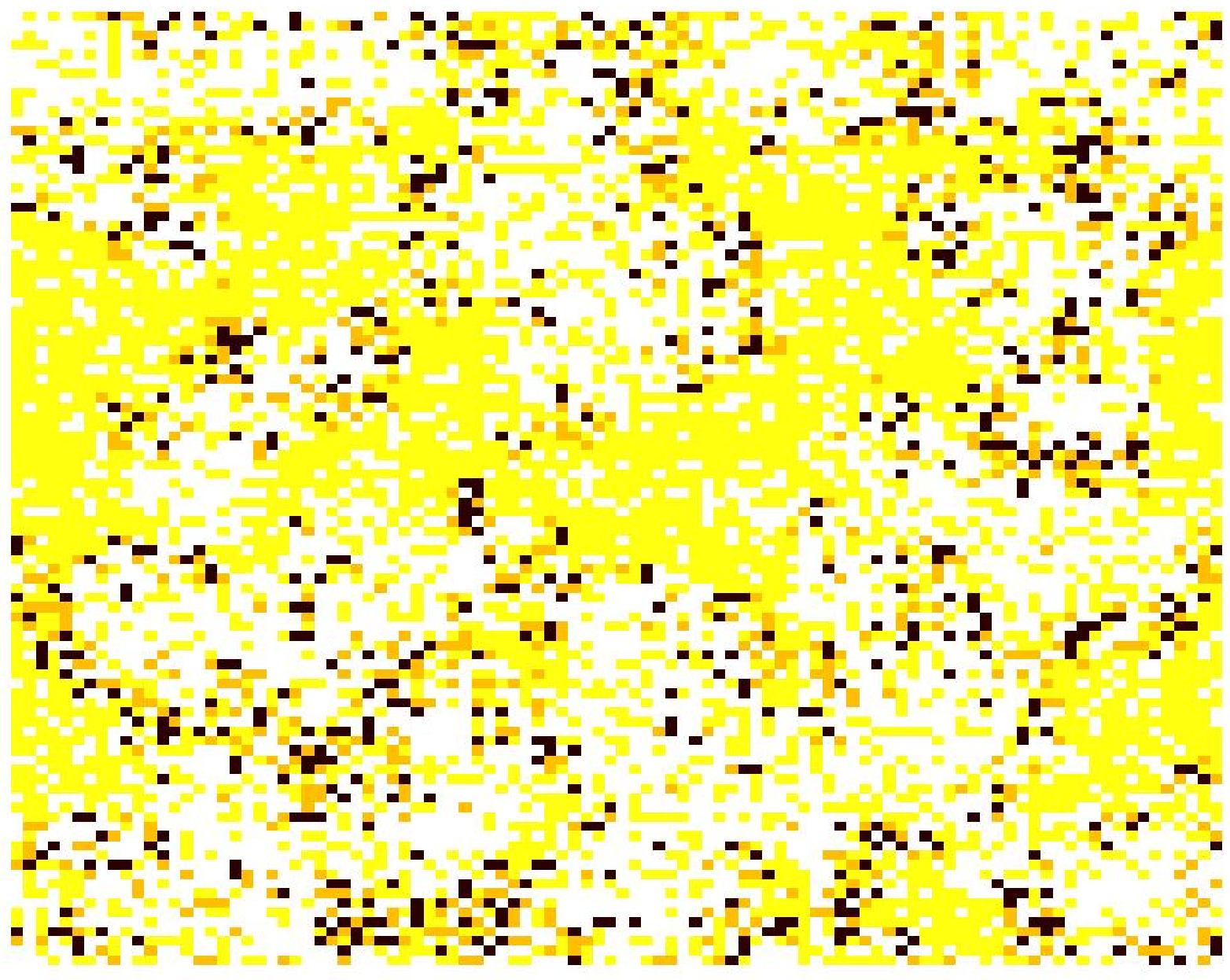}}

 \caption{In the above figures the space dispersion are
shown in details with parameters values $\alpha=\frac{1}{8},
\beta=\frac{1}{5}, \mu=0.4, d=0.1332, \lambda=0.45$ and $J=10000$.
Note that a given site may be empty (white), occupied by a
susceptible (grey), or occupied by an exposed, or infected
individuals (black). (a) and (b) are for low global density
$H(0)=0.4$, (b) The same after 50th iteration of the previous. (c)
and (d) are for high global density $H(0)=0.85$, (d) The same
after the 50th iteration of the previous.}\label{myfigure6}
\end{minipage}
\begin{minipage}[b]{0.48\textwidth}
\subfigure[]{
\label{fig7:a}\includegraphics[width=0.45\textwidth]{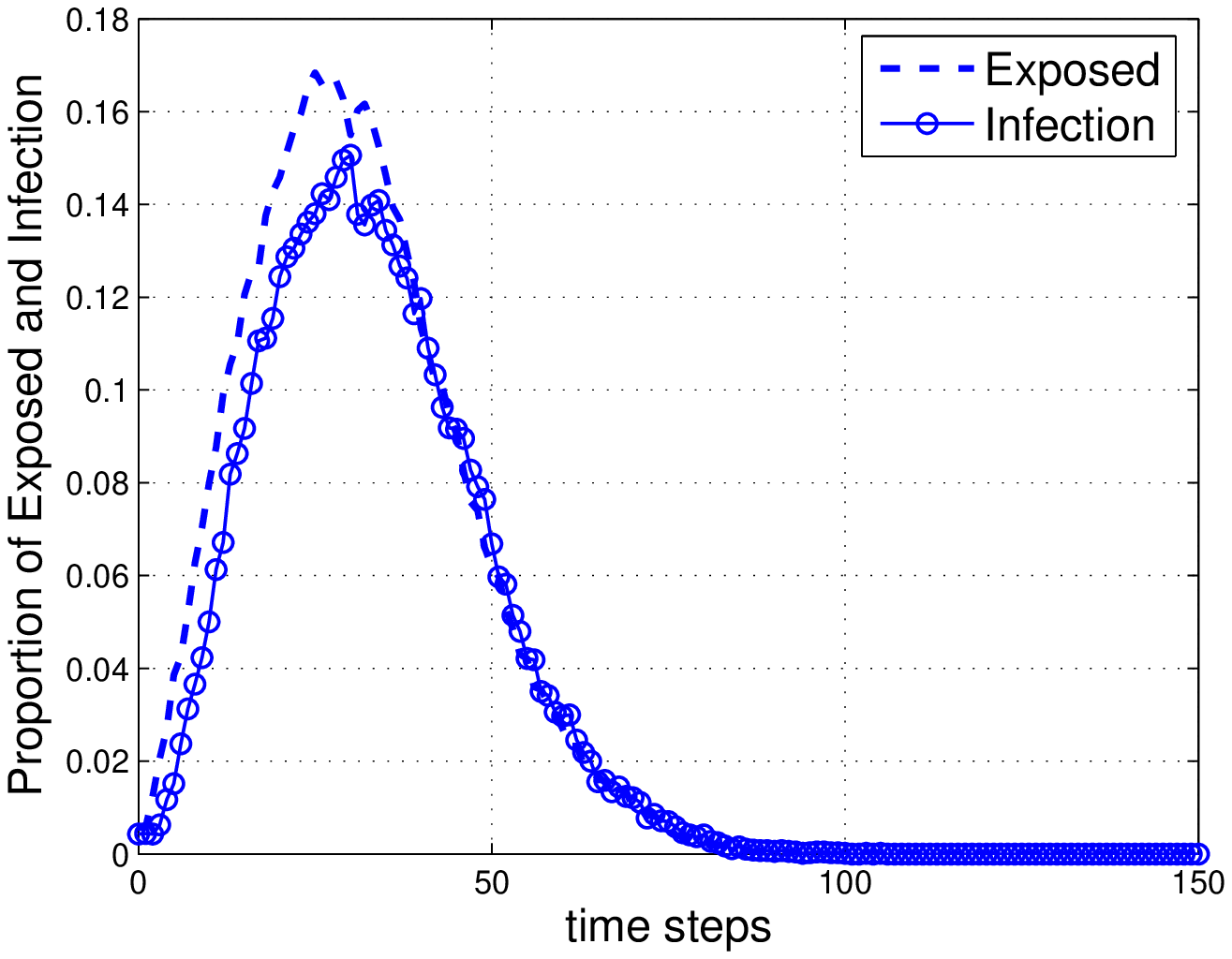}}
\subfigure[]{\label{fig7:a}\includegraphics[width=0.45\textwidth]{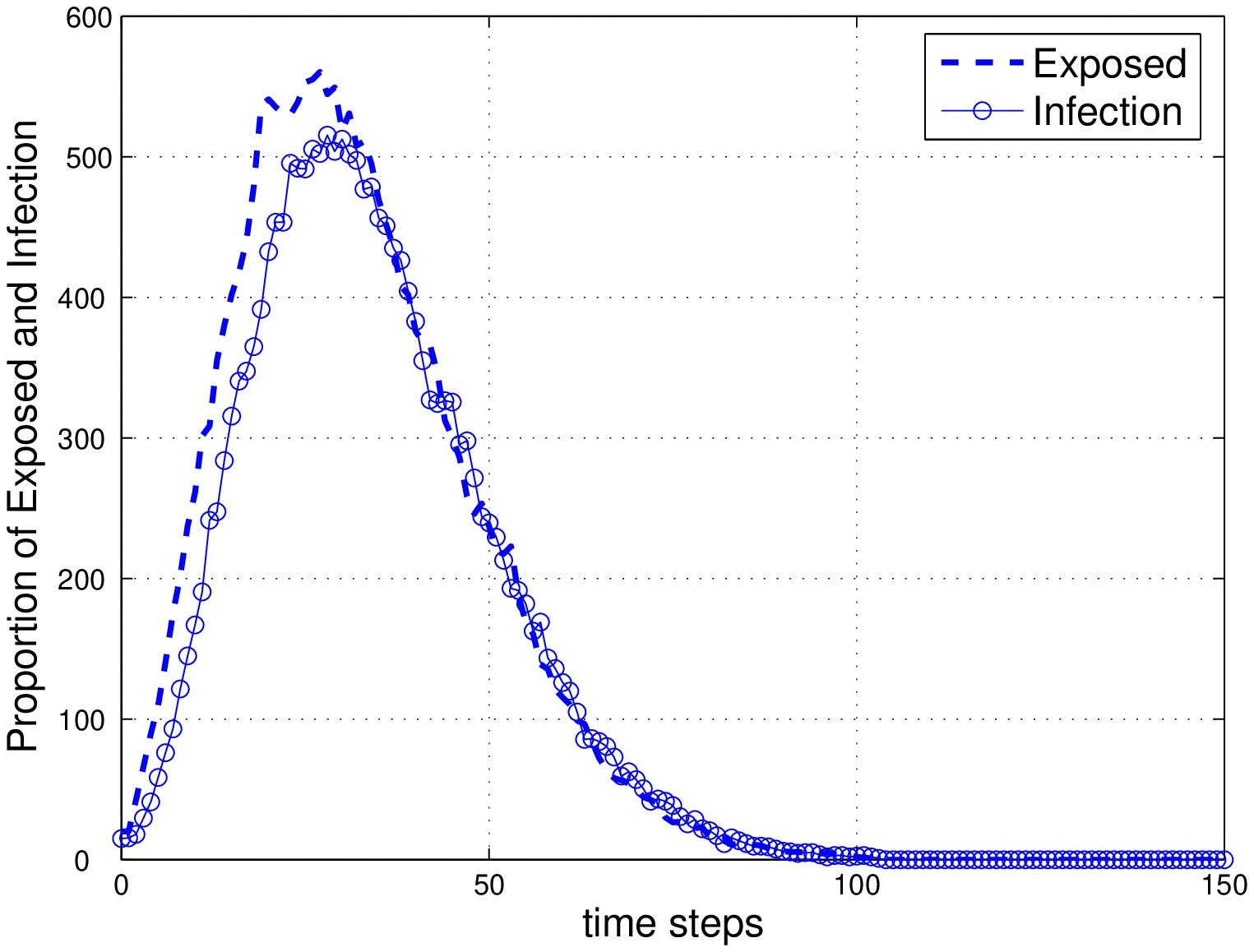}}

\caption{ In the above simulations we have taken the global
density $H=0.7$ and the mortality rate $d=0.1332$ (for  Hong Kong
city). The value of the other parameters are $\alpha=1/8$,
$\beta=1/5$, $ \mu=0.4$, $\lambda=0.5$. (a) The global frequency
of the exposed and infected with respect to time. (b) The numbers
of exposed and infected with respect to time.}\label{myfigure7}
\end{minipage}

 In the \autoref{myfigure6} the white, shaded and black
portion represent the empty site of susceptible, exposed and
infected individuals respectively. In \autoref{myfigure6}(a)
average density of the host is 0.4 and in \autoref{myfigure6}(c)
it is 0.85 at initial time t=0. We have got the
\autoref{myfigure6}(b) and \autoref{myfigure6}(d) at t=50 with the
same parameters values as in \autoref{myfigure6}(a) and
\autoref{myfigure6}(c) respectively.  The \autoref{myfigure7} is
the most important figure because it is  an example of simulation
for the SARS disease spreading  at China in 2003, where the global
density is equals to 0.7 (i.e $H=0.7$). The number of the recovery
of SARS is so small that we can ignore it, so the state of the
SARS is processing as the \autoref{myfigure04} shows. The detail
rule to the CA model is similar to \cite{Knu19}. The
\autoref{myfigure8} is
 collect from the papers of Steven et. al \cite{Knu9} where they have studied
about the SARS disease in Hong Kong city. Our \autoref{myfigure7}
is similar to their figure \autoref{myfigure8} but not same.
Similar works have performed by Lipsitch et. al ~\cite{Knu10}.
Therefore our results are supported by their works.\\

\begin{minipage}[b]{0.48\textwidth}
\subfigure[]{
\label{fig8:a}\includegraphics[width=0.45\textwidth]{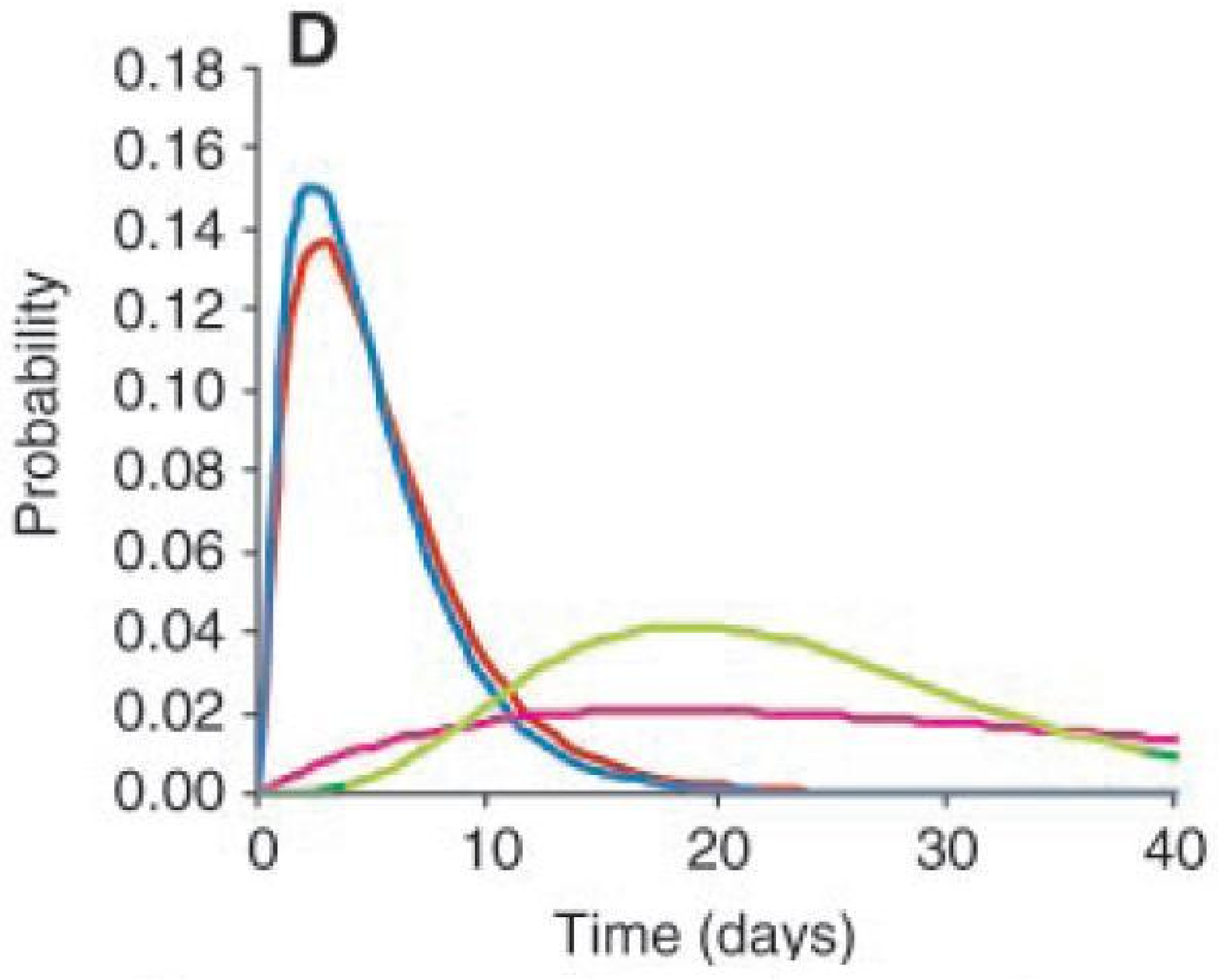}}
\subfigure[]{\label{fig8:b}\includegraphics[width=0.45\textwidth]{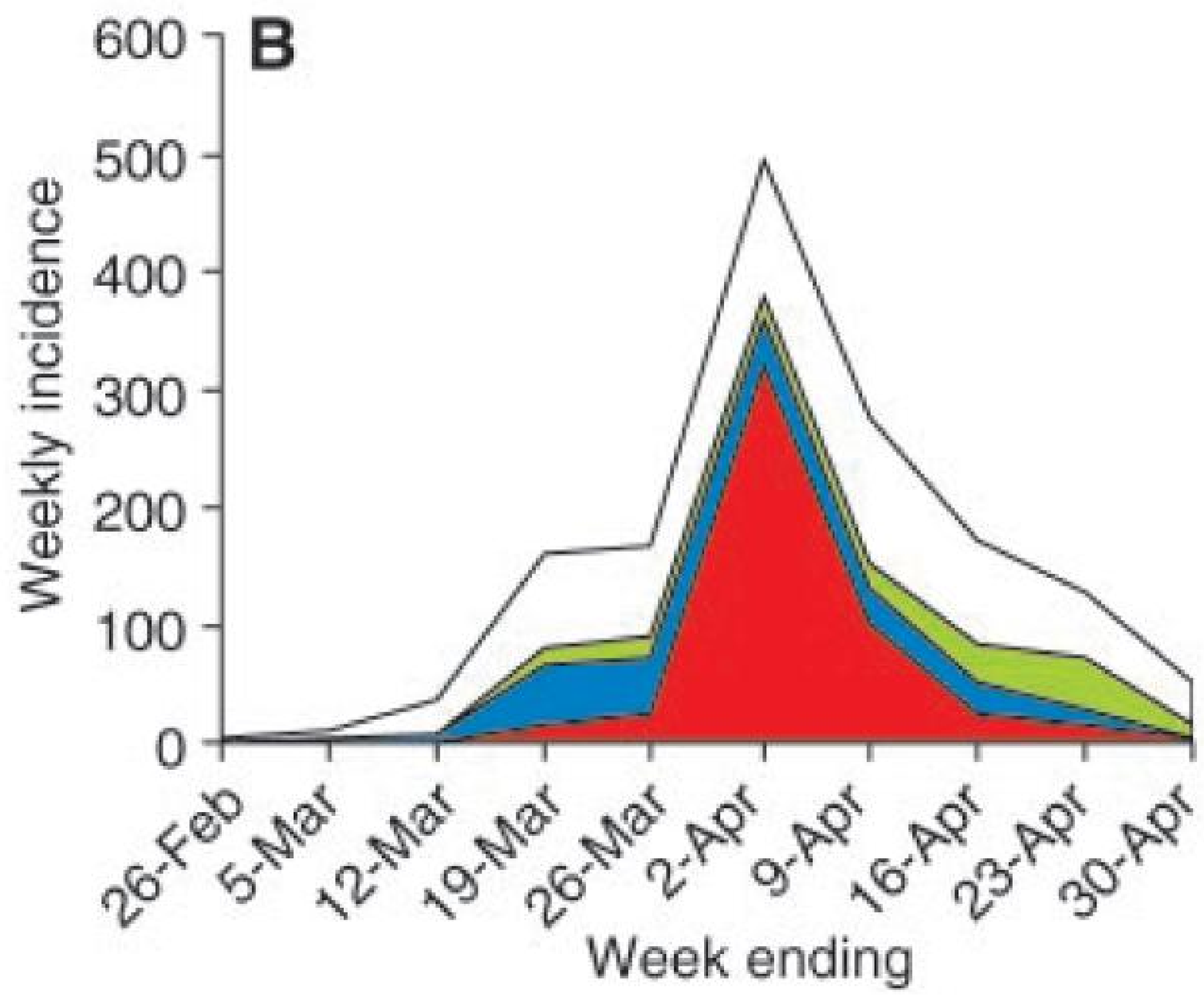}}

\caption{(a) Distributions for the waiting times of the
compartments of the stochastic model shown in see~\cite{Knu9}.
Distributions shown are estimates for the start of the epidemic.
The onset-to-hospitalization distribution changes during the
epidemic as a result of more rapid hospital admission. (b) Weekly
incidence (by time of hospital admission) with the color coding
used in (a).}\label{myfigure8}
\end{minipage}

\section{Conclusion and discussion}

In this article we have proposed and analyzed an epidemic model by
using cellular automata theory. We have considered here the
dispersion of the individuals with infectious force in latent
period. We have also used the standard $\textbf{Moore
neighborhood}$. The readers can compare the another methods of the
approximation to spatio-temporal models for epidemic with local
spread, such as PA, SA, HPA in~\cite{Knu20}.

 We have made some approximation which are very logical
for the system. The analytical findings and as well as the
numerical simulations suggest us that the frequency of exposed and
infected host increase with the increasing of heterogeneity
(density) and the size of neighborhood (See the figure 6, figure 7
and figure 8) .

 It has been observed that to spread the infection, the
epidemiological threshold $\mathcal{R}_c>1$. This Mean-Field
persister threshold ($R_{c}$) property is to some extent related
to the ``basic reproductive ratio $R_{0}$" of the epidemic theory,
see \cite{AM}.

 Generally, the effect of dispersion factor depends on
the size of neighborhood and global density of host. The
\autoref{myfigure6} have been performed with different values of
the global density of the host as the heterogeneity is an
important factor for epidemic models.

\section{Acknowledgments}

This work is supported by the National Natural Science Foundation
of China under Grant No 10471040 and part of this work have been
done at that time when author M. Haque is visiting North
University of China.

\newpage 


\begin{thebibliography}{22}
\expandafter\ifx\csname
natexlab\endcsname\relax\def\natexlab#1{#1}\fi
\expandafter\ifx\csname bibnamefont\endcsname\relax
  \def\bibnamefont#1{#1}\fi
\expandafter\ifx\csname bibfnamefont\endcsname\relax
  \def\bibfnamefont#1{#1}\fi
\expandafter\ifx\csname citenamefont\endcsname\relax
  \def\citenamefont#1{#1}\fi
\expandafter\ifx\csname url\endcsname\relax
  \def\url#1{\texttt{#1}}\fi
\expandafter\ifx\csname
urlprefix\endcsname\relax\def\urlprefix{URL }\fi
\providecommand{\bibinfo}[2]{#2}
\providecommand{\eprint}[2][]{\url{#2}}

\bibitem[{\citenamefont{Caracoa et~al.}(1998)\citenamefont{Caracoa, Duryea, adn
  William~Maniatty, and Szymanski}}]{Knu1}
\bibinfo{author}{\bibfnamefont{T.}~\bibnamefont{Caracoa}},
  \bibinfo{author}{\bibfnamefont{M.}~\bibnamefont{Duryea}},
  \bibinfo{author}{\bibfnamefont{G.~G.} \bibnamefont{adn William~Maniatty}},
  \bibnamefont{and} \bibinfo{author}{\bibfnamefont{B.~K.}
  \bibnamefont{Szymanski}}, \bibinfo{journal}{J. Theor. Biol.}
  \textbf{\bibinfo{volume}{351-61}}, \bibinfo{pages}{351}
  (\bibinfo{year}{1998}).

\bibitem[{\citenamefont{Fuk\'{s} and Lawniczak}(2001)}]{Knu14}
\bibinfo{author}{\bibfnamefont{H.}~\bibnamefont{Fuk\'{s}}} \bibnamefont{and}
  \bibinfo{author}{\bibfnamefont{A.~T.} \bibnamefont{Lawniczak}},
  \bibinfo{journal}{Discrete Dynamics in Nature and Society}
  \textbf{\bibinfo{volume}{6}}, \bibinfo{pages}{191} (\bibinfo{year}{2001}).

\bibitem[{\citenamefont{Mollison}(1991)}]{Knu2}
\bibinfo{author}{\bibfnamefont{D.}~\bibnamefont{Mollison}},
  \bibinfo{journal}{Mathematical Biosciences} \textbf{\bibinfo{volume}{107}},
  \bibinfo{pages}{255} (\bibinfo{year}{1991}).

\bibitem[{\citenamefont{Anderson et~al.}(1982)\citenamefont{Anderson, Gordon,
  Crawley, and Hassell}}]{Knu3}
\bibinfo{author}{\bibfnamefont{R.}~\bibnamefont{Anderson}},
  \bibinfo{author}{\bibfnamefont{D.}~\bibnamefont{Gordon}},
  \bibinfo{author}{\bibfnamefont{M.}~\bibnamefont{Crawley}}, \bibnamefont{and}
  \bibinfo{author}{\bibfnamefont{M.}~\bibnamefont{Hassell}},
  \bibinfo{journal}{Nature} \textbf{\bibinfo{volume}{296}},
  \bibinfo{pages}{245} (\bibinfo{year}{1982}).

\bibitem[{\citenamefont{Hutchings}(1988)}]{Knu4}
\bibinfo{author}{\bibfnamefont{M.}~\bibnamefont{Hutchings}},
  \emph{\bibinfo{title}{Plant population biology}}
  (\bibinfo{publisher}{Blackwell Scientific Publications (28th Symposium of the
  British Ecological Society)}, \bibinfo{year}{1988}).

\bibitem[{\citenamefont{Harcourt}(1961)}]{Knu5}
\bibinfo{author}{\bibfnamefont{D.}~\bibnamefont{Harcourt}},
  \bibinfo{journal}{Can. Ent.} \textbf{\bibinfo{volume}{93}},
  \bibinfo{pages}{945} (\bibinfo{year}{1961}).

\bibitem[{\citenamefont{Kobayashi}(1965)}]{Knu6}
\bibinfo{author}{\bibfnamefont{S.}~\bibnamefont{Kobayashi}},
  \bibinfo{journal}{Res. Popul. Ecol.} \textbf{\bibinfo{volume}{7}},
  \bibinfo{pages}{109} (\bibinfo{year}{1965}).

\bibitem[{\citenamefont{Duryea et~al.}(19999)\citenamefont{Duryea, Caraco,
  Garder, Maniatty, and Szymanski}}]{Knu7}
\bibinfo{author}{\bibfnamefont{M.}~\bibnamefont{Duryea}},
  \bibinfo{author}{\bibfnamefont{T.}~\bibnamefont{Caraco}},
  \bibinfo{author}{\bibfnamefont{G.}~\bibnamefont{Garder}},
  \bibinfo{author}{\bibfnamefont{W.}~\bibnamefont{Maniatty}}, \bibnamefont{and}
  \bibinfo{author}{\bibfnamefont{B.}~\bibnamefont{Szymanski}},
  \bibinfo{journal}{Physica D} \textbf{\bibinfo{volume}{132}},
  \bibinfo{pages}{511} (\bibinfo{year}{19999}).

\bibitem[{\citenamefont{Ahmed and Agiza}(1998)}]{Knu15}
\bibinfo{author}{\bibfnamefont{E.}~\bibnamefont{Ahmed}} \bibnamefont{and}
  \bibinfo{author}{\bibfnamefont{H.}~\bibnamefont{Agiza}},
  \bibinfo{journal}{Physica A} \textbf{\bibinfo{volume}{253}},
  \bibinfo{pages}{347} (\bibinfo{year}{1998}).

\bibitem[{\citenamefont{Keeling}(1999)}]{keel}
\bibinfo{author}{\bibfnamefont{M.}~\bibnamefont{Keeling}},
  \bibinfo{journal}{Proc. Roy. Soc. Lond. B} \textbf{\bibinfo{volume}{266}},
  \bibinfo{pages}{859} (\bibinfo{year}{1999}).

\bibitem[{\citenamefont{San-Ling}(2001)}]{Knu17}
\bibinfo{author}{\bibfnamefont{Y.}~\bibnamefont{San-Ling}},
  \bibinfo{journal}{J. Biol. Math.} \textbf{\bibinfo{volume}{16}},
  \bibinfo{pages}{392} (\bibinfo{year}{2001}).

\bibitem[{\citenamefont{Guihua and Zhen}(2005)}]{Knu18}
\bibinfo{author}{\bibfnamefont{L.}~\bibnamefont{Guihua}} \bibnamefont{and}
  \bibinfo{author}{\bibfnamefont{J.}~\bibnamefont{Zhen}},
  \bibinfo{journal}{Chaos, Solitons and Fractals}
  \textbf{\bibinfo{volume}{25}}, \bibinfo{pages}{1177} (\bibinfo{year}{2005}).

\bibitem[{\citenamefont{Hiebeler}(1997)}]{Knu8}
\bibinfo{author}{\bibfnamefont{D.}~\bibnamefont{Hiebeler}},
  \bibinfo{journal}{J. Theor. Biol.} \textbf{\bibinfo{volume}{187}},
  \bibinfo{pages}{307} (\bibinfo{year}{1997}).

\bibitem[{\citenamefont{Diekman and Heesterbeek}(2000)}]{ODk}
\bibinfo{author}{\bibfnamefont{O.}~\bibnamefont{Diekman}} \bibnamefont{and}
  \bibinfo{author}{\bibfnamefont{J.}~\bibnamefont{Heesterbeek}},
  \emph{\bibinfo{title}{Mathematical Epidemiology of Infectious Disease-Model
  Building, Aalysis and Interpretatio}}, Wiley series in mathematical and
  Computational Biology (\bibinfo{publisher}{John Wiley and Son, Ltd},
  \bibinfo{year}{2000}).

\bibitem[{\citenamefont{Laselle}(1993)}]{Knu12}
\bibinfo{author}{\bibfnamefont{J.}~\bibnamefont{Laselle}},
  \emph{\bibinfo{title}{The stability of dynamical system}}
  (\bibinfo{publisher}{Hmilton Borlin press}, \bibinfo{year}{1993}).

\bibitem[{\citenamefont{Jury}(1974)}]{Knu13}
\bibinfo{author}{\bibfnamefont{E.}~\bibnamefont{Jury}},
  \emph{\bibinfo{title}{Inners and stability of dynamic systems}}
  (\bibinfo{publisher}{Wiley}, \bibinfo{address}{New york},
  \bibinfo{year}{1974}).

\bibitem[{\citenamefont{Ellner}(2001)}]{Knu16}
\bibinfo{author}{\bibfnamefont{S.~P.} \bibnamefont{Ellner}},
  \bibinfo{journal}{J. Theor. Biol.} \textbf{\bibinfo{volume}{210}},
  \bibinfo{pages}{435} (\bibinfo{year}{2001}).

\bibitem[{\citenamefont{Quanxing and Zhen}(2005)}]{Knu19}
\bibinfo{author}{\bibfnamefont{L.}~\bibnamefont{Quanxing}} \bibnamefont{and}
  \bibinfo{author}{\bibfnamefont{J.}~\bibnamefont{Zhen}},
  \bibinfo{journal}{Chinese Physics} \textbf{\bibinfo{volume}{15}},
  \bibinfo{pages}{1370} (\bibinfo{year}{2005}).

\bibitem[{\citenamefont{Riley et~al.}(2003)\citenamefont{Riley, Fraser, and
  Donnelly}}]{Knu9}
\bibinfo{author}{\bibfnamefont{S.}~\bibnamefont{Riley}},
  \bibinfo{author}{\bibfnamefont{C.}~\bibnamefont{Fraser}}, \bibnamefont{and}
  \bibinfo{author}{\bibfnamefont{C.~A.} \bibnamefont{Donnelly}},
  \textbf{\bibinfo{volume}{300}}, \bibinfo{pages}{1966} (\bibinfo{year}{2003}).

\bibitem[{\citenamefont{Lipsitch et~al.}(2003)\citenamefont{Lipsitch, Cohen,
  and Cooper}}]{Knu10}
\bibinfo{author}{\bibfnamefont{M.}~\bibnamefont{Lipsitch}},
  \bibinfo{author}{\bibfnamefont{T.}~\bibnamefont{Cohen}}, \bibnamefont{and}
  \bibinfo{author}{\bibfnamefont{B.}~\bibnamefont{Cooper}},
  \bibinfo{journal}{Science} \textbf{\bibinfo{volume}{300}},
  \bibinfo{pages}{1966} (\bibinfo{year}{2003}).

\bibitem[{\citenamefont{Filope and Gibson}(2001)}]{Knu20}
\bibinfo{author}{\bibfnamefont{J.}~\bibnamefont{Filope}} \bibnamefont{and}
  \bibinfo{author}{\bibfnamefont{G.}~\bibnamefont{Gibson}},
  \bibinfo{journal}{Bulletin of Mathematical Biology}
  \textbf{\bibinfo{volume}{63}}, \bibinfo{pages}{603} (\bibinfo{year}{2001}).

\bibitem[{\citenamefont{Anderson and May}(1981)}]{AM}
\bibinfo{author}{\bibfnamefont{R.}~\bibnamefont{Anderson}} \bibnamefont{and}
  \bibinfo{author}{\bibfnamefont{R.}~\bibnamefont{May}},
  \bibinfo{journal}{Proc. Roy. Soc. Lond. B} \textbf{\bibinfo{volume}{291}},
  \bibinfo{pages}{451} (\bibinfo{year}{1981}).

\end{thebibliography}
\end{document}